\documentclass[12pt]{article}
\usepackage{latexsym,amssymb}
\usepackage[american]{babel}
\makeatletter
\renewcommand{\@evenhead}{\raisebox{0pt}[\headheight][0pt]{\vbox{\hbox
to \textwidth{\thepage\hfil\strut\textsc{\leftmark}}\hrule}}}
\renewcommand{\@oddhead}{\raisebox{0pt}[\headheight][0pt]{\vbox{\hbox
to \textwidth{\textsc{\rightmark}\hfil\strut\thepage}\hrule}}}
\makeatother

\def\II{{\mathbb I}}
\def\RR{{\mathbb R}}
\def\CC{{\mathbb C}}

\def\ZZ{{\mathbb Z}}
\def\NN{{\mathbb N}}

\def\Alt{\mathop{\rm Alt}\nolimits}

\def\Id{{\rm Id\,}}

\def\tr{{\rm tr\,}}
\def\Tr{{\rm Tr\,}}

\def\det{{\rm det\,}}

\def\Cliff{{\rm Cliff}}
\def\Span{{\rm Span}}
\def\Pin{{\rm Pin}}
\def\Spin{{\rm Spin}}
\def\Aut{{\rm Aut}}
\def\Mat{{\rm Mat}}
\def\End{{\rm End}}
\def\for{\qquad {\rm for} \qquad}
\def\and{\qquad {\rm and} \qquad}
\def\Id{{\rm Id}}

\def\mod{{\rm mod\,}}
\def\vol{{\rm vol\,}}

\def\Ker{{\rm \,Ker\,}}

\def\be{\begin{equation}}
\def\ee{\end{equation}}
\def\bea{\begin{eqnarray}}
\def\eea{\end{eqnarray}}

\def\sideremark#1{\ifvmode\leavevmode\fi\vadjust{\vbox to0pt
{\vss\hbox to 0pt{\hskip\hsize\hskip1em
\vbox{\hsize2cm\tiny\raggedright\pretolerance10000
\noindent #1\hfill}\hss}\vbox to8pt{\vfil}\vss}}}

\begin{document}


\begin{titlepage}

\null
\vskip-2cm 
\hfill
\begin{minipage}{5cm}
\par\hrulefill\par\vskip-4truemm\par\hrulefill
\par\vskip2mm\par
{{\large\sc New Mexico Tech \\[12pt]
{\rm (January 2005)}}}
\par\hrulefill\par\vskip-4truemm\par\hrulefill
\end{minipage}
\bigskip 
\bigskip

\par
\hfill
\vfill

\centerline{\LARGE\bf Dirac Operator in Matrix Geometry} 
\bigskip
\bigskip 
\centerline{\Large\bf Ivan G. Avramidi} 
\bigskip 
\centerline{\it Department of Mathematics} 
\centerline{\it New Mexico Institute of Mining and Technology} 
\centerline{\it Socorro, NM 87801, USA} 
\centerline{\it E-mail: iavramid@nmt.edu} 
\bigskip
\bigskip
\centerline{\sc 
Dedicated to the Memory of Vladimir Fock and Dmitri Ivanenko}
\medskip 
\vfill 
{\narrower 

We review the construction of the Dirac operator and its properties in
Riemannian geometry and show how the asymptotic expansion of the
trace of the heat kernel determines the spectral invariants of the Dirac
operator and its index. We also point out that the Einstein-Hilbert
functional can be obtained as a  linear combination of the first two
spectral invariants of the Dirac operator. Next, we report on our
previous attempts to generalize the notion of the Dirac operator to the
case of Matrix Geometry, when instead of a Riemannian metric there is a
matrix valued self-adjoint symmetric two-tensor that plays a role of a
``non-commutative'' metric. We construct invariant first-order and
second-order self-adjoint elliptic partial differential operators, that
can be called ``non-commutative'' Dirac operator and non-commutative
Laplace operator. We construct the corresponding heat kernel for the
non-commutative Laplace type operator and compute its first two spectral
invariants. A linear combination of these two spectral invariants gives
a  functional that can be considered as a non-commutative generalization
of the Einstein-Hilbert action.

}

\end{titlepage}


\section{Introduction}

Dirac operator was discovered by Dirac in 1928 as a ``square root'' of 
the D'Alambert operator in a flat Minkowskian space in an attempt to
develop a relativistic theory of the electron. In 1929, almost
immediately after  Dirac's paper, Fock and Ivanenko
\cite{fockivanenko29a, fockivanenko29b} showed how to generalize the
Dirac's equation for the  case of General Relativity. Fock  completed
the geometrization of the theory of spinors on  Riemannian manifolds in
\cite{fock29a, fock29b,fock29c}. This development was purely local. 
Only much later, at the end of forties in a global setting it was
understood that there are topological obstructions to the existence of
the spinor structure and spinor fields and Dirac operators cannot be
introduced on every Riemannian manifold (see, for example,
\cite{lawson89, berline92, friedrich00}). 

The construction of a square root of the Laplacian naturally leads to 
the study of complex representations of the Clifford algebra. The
spinors are then introduced as the elements of the corresponding vector
space. It turns out that there are no non-trivial representations of the
orthogonal group in the vector space of spinors compatible with Clifford
multiplication. Therefore, spinors on a Riemannian manifold  cannot be
introduced as sections of a vector bundle associated with the frame
bundle of the manifold. Instead of the special orthogonal group   one
considers its double covering group, so called spin group. Contrary to
the orthogonal group the spin group has a representation in the vector
space of spinors compatible with Clifford multiplication. Now, if the
frame bundle of the manifold allows a reduction to the spin group,
then the manifold is said to admit a spin structure and one can define
the spinor bundle, which is a vector bundle associated with this
reduction via the representation of the spin group. The spinors are the
sections of the spinor bundle.

The spinor bundle inherits a connection from the canonical Levi-Civita
connection, which enables one to define the Dirac operator. The Dirac
operator on a spinor bundle is now called the Dirac operator in the
narrow sense, while the general (also called twisted)  Dirac operator
(or Dirac type operator)  is any self-adjoint first-order operator whose
square is equal to the D'Alambert operator (or the Laplace operator). 

The Dirac operator in Riemannian geometry is defined with the help of a
Riemannian metric and the spin connection.  Following the ideas of our 
papers \cite{avramidi03, avramidi04a, avramidi04b}  we are going to
generalize this formalism to the case of Matrix Geometry, when instead
of a single Riemannian metric there is a matrix-valued symmetric
2-tensor, which we call a ``non-commutative metric''.  Matrix Geometry  
is motivated by the relativistic interpretation of gauge theories and is
intimately related to Finsler geometry (rather a collection of Finsler
geometries) (see \cite{avramidi03, avramidi04a, avramidi04b}). In the
present paper we will not discuss the origin of  Matrix Geometry, but
simply assume the existense of such a structure.  We will not be
concerned about the global issues as well.  We will simply assume that
there are no topological obstructions to all the structures introduced
below.  Our ``non-commutative'' Dirac operator is  a first-order 
elliptic partial differential operator such that its square is a
second-order self-adjoint elliptic operator with positive definite
leading symbol (not necessarily of Laplace type).

The outline of the paper is as follows.  First, we review the
construction of the Dirac operator in Riemannian geometry. In section
2.1 we introduce the Clifford algebra and describe its properties. In
section 2.2 we introduce the spin group and show that it is a double
cover of the orthogonal group. In section 2.3 the spin representation of
the spin group is introduced and the spinors are defined. In section 2.4
the derivation of the spin connection and its curvature is described. In
section 2.5 we define the Dirac operator and its index.  In section 2.6
we introduce the heat kernel. It is explained how the asymptotic
expansion of the trace of the heat kernel generates the spectral
invariants of the Dirac operator, in particular, its index.

In section 3.1 we introduce non-commutative (or matrix) generalization
of the Riemannian metric and the Dirac matrices as a deformation of the
commutative limit. In section 3.2  we  introduce non-commutative
versions of the Hodge star operator acting on space of matrix valued
$p$-forms. In section 3.3  the relation of the matrix geometry to
Finsler geometry is explored.  In section 3.4 we promote the vector
spaces introduced above to vector bundles and  define the corresponding
Hilbert spaces. Since we do not have a Riemannian volume element, we
work here not with tensors but rather with densities of various weight.
In section 3.5 we develop an exterior calculus for matrix-valued
densities and in section 3.6 we introduce a ``non-commutative
connection''. Sections 3.7 and 3.8 discuss the construction of the
non-commutative versions of the Dirac operator and the Laplacian. In
sections 3.9 and 3.10 we discuss the spectral asymptotics of the
operators introduced before and compute the first two spectral
invariants.

In section 4 we construct a non-commutative deformation of the 
Einstein-Hilbert functional.


\section{Dirac Operators in Riemannian Geometry}

\subsection{Clifford Algebra}

Let $M$ be an $n$-dimensional manifold, $x$ be a point in $M$ and
$V=T_xM$ be the tangent space at the point $x$ (which is isomorphic to
$\RR^n$) equipped with a positive-definite scalar product $\langle\ , 
\ \rangle$. 
Let $m=\left[{n\over 2}\right]$ so that for even dimension $n=2m$ and
for odd dimension $n=2m+1$. Let also $N=2^m$. 
The real Clifford algebra
$\Cliff(n)$ is the  universal associative algebra with unit generated
multiplicatively by the range $\gamma(V)$ of a linear map
\be
\gamma: V\to\Cliff(n)
\ee
satisfying
\be
\gamma(u)\gamma(v)+\gamma(v)\gamma(u)=2\langle u,v\rangle\II\,,
\ee
where $\II$ is the unit in the algebra.
In particular,
\be
\gamma(u)\gamma(u)=|u|^2\II\,,
\ee
where $|u|^2=\langle u,u\rangle$, 
and, therefore, for any unit vector $u$, $|u|=1$,
the element $\gamma(u)$ is invertible and
\be
[\gamma(u)]^{-1}=\gamma(u)\,.
\ee

Let $\otimes V$ be the tensor algebra
\be
\otimes V=\bigoplus_{k=0}^\infty 
\otimes^k V\,,
\ee
where $\otimes^0 V=\RR$, and ${\cal I}$ be the ideal generated by the set  
$$
\{u\otimes v+v\otimes u-2\langle u,v\rangle \;|\; u,v\in V\}\,.
$$
Then the Clifford algebra can be identified with the quotient
\be
\Cliff(n)=\otimes V/{\cal I}\,.
\ee
The corresponding complex Clifford algebra 
is obtained by tensoring the real algebra with the complex numbers
\be
\Cliff_\CC(n)=\Cliff(n)\otimes_{\RR}\CC\,.
\ee 

The tensor algebra $\otimes V$ has a natural $\NN$-grading, which
after reduction $\mod 2$ leads to a natural $\ZZ_2$-grading
of the Clifford algebra
\be
\Cliff(n)=\Cliff_+(n)\oplus\Cliff_-(n)\,,
\ee
where $\Cliff_+(n)$ and $\Cliff_-(n)$ are the even and the odd
parts respectively consisting of the sums
of products of even and odd number of elements from $\gamma(V)$
with
\be
\Cliff_k(n)\Cliff_j(n)\subset \Cliff_{jk}(n)\,,
\ee
where $k,j=\pm 1$.
Therefore, the even part $\Cliff_+(n)$ is a subalgebra of the Clifford algebra
$\Cliff(n)$.

Let $F_0(n)=\RR$ and for $1\le k\le n$
\be
F_k(n)=\Span \{\gamma(u_1)\cdots\gamma(u_j) \;|\; 1\le j\le k, \ u_i\in V\}
\ee
be the subspace of $\Cliff(n)$ consisting of the sums
of the products of at most $k$ elements from $\gamma(V)$.
Then the Clifford algebra has a natural increasing
filtration 
\be
F_0(n)\subset F_1(n)\subset \cdots \subset F_n(n)=\Cliff(n)
\ee
such that
\be
F_j(n)F_k(n)\subset F_{k+j}(n)\,,
\ee
where, by definition $F_j(n)=F_n(n)$ if $j>n$.

Further, let
\be
C_0(n)=\RR\,,\qquad 
C_k(n)=F_k(n)/F_{k-1}(n), \qquad 1\le k\le n\,.
\ee
Then
\be
F_{2k}(n)=\bigoplus_{j=0}^k C_{2k-2j}(n)  
\ee
and
\be
F_{2k+1}(n)=\bigoplus_{j=0}^k C_{2k-2j+1}(n)\,.  
\ee
The space $C_k(n)$ 
is isomorphic to $\wedge^k V$ and the Clifford algebra is a graded algebra, 
which, as a vector space, has the form
\be
\Cliff(n)=\bigoplus_{k=0}^n C_k(n)\,,
\ee
with
\be
C_j(n)C_k(n)=\bigoplus_{0\le 2l\le k+j} C_{k+j-2l}(n)\,,
\ee
and is naturally isomorphic to the exterior algebra $\wedge V$
\be
\wedge V=\bigoplus_{k=0}^{n} 
\wedge^k V\,,
\ee
where $\wedge^0 V=\RR$.

Therefore the map $\gamma: V\to \Cliff(n)$ can be extended to an
isomorphism 
\be
\gamma: \wedge V\to\Cliff(n)
\ee
of the exterior algebra and the Clifford
algebra such that 
\be
\gamma(1)=\II\,.
\ee
We also have
\be
\Cliff_+(n)=\bigoplus_{0\le 2k\le n}^{}C_{2k}(n)\,,\qquad
\Cliff_-(n)=\bigoplus_{0\le 2k+1\le n}^{}C_{2k+1}(n)\,.
\ee

There are natural projections
\be
{\rm Pr}_k: \Cliff(n)\to C_k(n)\,.
\ee
The projection onto the unit element
\be
{\rm Pr}_0: \Cliff(n)\to C_0(n)=\RR\,
\ee
defines a natural linear functional on the Clifford algebra, which satisfies
a very important property
\be
{\rm Pr}_0(AB)={\rm Pr}_0(BA)\,,
\ee
and a normalization condition
\be
{\rm Pr}_0(\II)=1\,.
\ee
One can conclude from this that
\be
{\rm Pr}_0 C_k(n)=0 \qquad {\rm for} \ k\ne 0\,,
\ee
and
\be
{\rm Pr}_0 F_{2k+1}(n)=0\,.
\ee

There is a natural involution 
\be
\alpha: \Cliff(n)\to \Cliff(n)\,,
\ee
such that
\be
\alpha^2=\Id\, \qquad {\rm and} \qquad
\alpha(AB)=\alpha(A)\alpha(B)\,
\ee
defined by
\be
\alpha[\gamma(u_1)\cdots\gamma(u_k)]=(-1)^k\gamma(u_1)\cdots\gamma(u_k)\,.
\ee
Then
\be
\alpha(A)=(-1)^kA \qquad {\rm for} \ A\in C_k(n)\,.
\ee
and
\be
\alpha(A)=\varepsilon(A) A 
\ee
where
\be
\varepsilon(A)=\pm 1 \for 
A\in \Cliff_{\pm}(n)\,
\ee
is the parity of the element $A$.

There is a natural transpose on the tensor algebra $\otimes V$ defined by
\be
u_1\otimes \cdots \otimes u_k \mapsto u_k\otimes \cdots\otimes u_1\,.
\ee
Since the ideal ${\cal I}$ 
is preserved under this action, there is a natural
linear anti-involution of the Clifford algebra (reversing map, or
transposition) 
\be
\tau: \Cliff(n)\to \Cliff(n)
\ee
such that
\be
\tau^2=\Id, \and \tau(AB)=\tau(B)\tau(A) \,
\ee
defined by
\be
\tau[\gamma(u_1)\cdots\gamma(u_k)]=\gamma(u_k)\cdots\gamma(u_1)\,.
\ee

The composition of the above maps defines another anti-involution 
(conjugation)
\be
*=\tau\circ\alpha: \Cliff(n)\to \Cliff(n)
\ee
such that
\be
*^2=\Id, \and (AB)^*=B^*A^*\,
\ee
by
\be
[\gamma(u_1)\cdots\gamma(u_k)]^*=(-1)^k\gamma(u_k)\cdots\gamma(u_1)\,.
\ee
Note that 
\be
A^*=\varepsilon(A)\tau(A)\,,
\ee
so that for the even Clifford subalgebra the anti-involutions $\tau$ and $*$
coincide.


The center of the Clifford algebra is one-dimensional in even dimension
$n=2m$ and two-dimensional in odd dimension $n=2m+1$.
More precisely,
\bea
{\cal Z}(\Cliff(2m))&=&C_0(n)\\[10pt]
{\cal Z}(\Cliff(2m+1))&=&C_0(2m+1)\oplus C_{2m+1}(2m+1)\,.
\eea
This simply means that the only elements that commute with all elements
of the Clifford algebra have the form $a\II$ in even dimension $n=2m$ and 
$a\II+b\gamma(e_1)\cdots \gamma(e_{2m+1})$ in odd dimension $n=2m+1$,
where $a,b$ are scalars and the vectors $\{e_a\}$ are
orthogonal to each other.

Let $\{e_a\}=\{e_1,\cdots,e_n\}$, where 
$a=1,\dots,n$, be an oriented orthonormal basis of
$V$, that is
\be
\langle e_a,e_b\rangle=\delta_{ab}\,,
\ee
where $\delta_{ab}$ is the Kronecker symbol.
We use small Latin letters running over $1,\dots, n$ to denote
vectors from the vector space $V$.
We also use the  standard summation convention to
sum over repeated indices. Such indices will be raised and lowered by the 
Euclidean metric (the Kronecker symbol $\delta_{ab}$). 
Then the elements
\be
\gamma_a=\gamma(e_a)
\ee
of the Clifford algebra 
satisfy the anti-commutation relations
\be
\gamma_a\gamma_b+\gamma_b\gamma_a=2\delta_{ab}\II\,.
\ee
Thus, $\gamma_a$ are involutions that anti-commute with each other
\bea
(\gamma_a)^2&=&\II, \qquad\\[5pt]
\gamma_a\gamma_b&=&-\gamma_b\gamma_a, \qquad {\rm for} \ a\ne b\,.
\eea
The Clifford algebra $\Cliff(n)$ 
is multiplicatively generated by the elements 
$\gamma_a$.

Let $S_k$ be the permutation group of integers $(1,\dots,k)$. The
signature ${\rm sgn}(\sigma)$ (or the sign, or the parity)  of a
permutation $\sigma\in S_k$ is defined to be $+1$ if $\sigma$ is even
and $-1$ if $\sigma$ is odd. The complete antisymmetrization of a tensor
$T_{a_1\cdots a_k}$ over the indices $a_1,\dots, a_k$, is denoted by the
square brackets, and is defined by
\be
T_{[a_1\cdots a_k]}={1\over k!}
\sum\limits_{\sigma\in S_k} {{\rm sgn}(\sigma)}
T_{a_{\sigma(1)}\cdots a_{\sigma(k)}}\,,
\ee
where the summation is taken over the $k!$ permutations 
of $(1,\dots,k)$.
Let us further define the anti-symmetrized products of $\gamma_a$
\be
\gamma_{a_1\cdots a_k}=\gamma_{[a_1}\cdots\gamma_{a_k]}\,.
\label{1.191}
\ee
Of course, these elements are completely anti-symmetric in all their
indices. They are non-zero only when all
indices are different. In this case
\be
\gamma_{a_1\cdots a_k}=\gamma_{a_1}\cdots \gamma_{a_k}\,.
\ee

Obviously, for $1\le k\le n$ we have
\be
C_k(n)=\Span\left\{\gamma_{a_1\cdots a_k}\;|\; 
\ \ 1\le a_j\le n, \ \ 1\le j\le k
\right\}\,,
\ee
and
\be
\Cliff(n)=\Span\{\II, \gamma_{a_1\cdots a_k}\;|\; \ \ 1\le a_j\le n, \ \ 
1\le j \le k, \ \ 1\le k\le n\}\,.
\ee
That is each element of the Clifford algebra 
$\Cliff(n)$ is a linear combination
of the elements $\gamma_{a_1\cdots a_k}$ with real coefficients.

The extension of the map $\gamma$ to the whole exterior algebra 
$\wedge V$ is defined by
\be
\gamma(1)=\II\,,\qquad 
\gamma(e_{a_1}\wedge\cdots\wedge e_{a_k})=\gamma_{a_1\cdots a_k}\,.
\ee
Therefore, the elements
\be
\II, \ \gamma_{a_1}, \ \gamma_{a_1a_2}, \ \dots, \ \gamma_{a_1\cdots a_{n}},
\ee
where $(1\le a_1<a_2<\cdots<a_{n}\le n)$ form a basis
in the vector space $\Cliff(V)$.
The number of elements in the basis is $2^n$.
Therefore, the Clifford algebra $\Cliff(n)$ has the dimension
\be
\dim\Cliff(n)=2^n.
\ee
The dimension of the even subalgebra $\Cliff_+(n)$ is equal to one half
of $\dim\Cliff(n)$, that is, $2^{n-1}$.

Let us define the chirality element $\Gamma\in \Cliff_\CC(n)$ by
\be
\Gamma=i^{n(n-1)/2}\gamma_{1\cdots n}
={i^{n(n-1)/2}\over n!}\varepsilon^{a_1\cdots a_{n}}\gamma_{a_1\cdots a_n}\,,
\label{c1}
\ee
where $\varepsilon^{a_1\cdots a_n}$ is the completely antisymmetric
Levi-Civita symbol normalized by $\varepsilon^{1\cdots n}=+1$. There is
an ambiguity of choosing the sign of the chirality operator $\Gamma$
corresponding to the choice of the orientation of the vector space $V$.
The chirality operator is an involution, that is
\be
\Gamma^2=\II
\ee
that anticommutes with all
$\gamma_a$ in even dimensions and commutes with all $\gamma_a$
in odd dimensions. That is,
\be
\Gamma \gamma_a=-\gamma_a\Gamma\,, \qquad  {\rm for \ even \ } n\,,
\ee
\be
\Gamma \gamma_a=\gamma_a\Gamma\,, \qquad  {\rm for \ odd \ } n\,.
\ee
Thus in odd dimension $\Gamma$ lies in the center of the Clifford algebra,
and in even dimension we have
\be
\Gamma\gamma_{a_1\cdots a_k}=(-1)^k\gamma_{a_1\cdots a_k}\Gamma
\qquad  {\rm for \ even \ } n\,.
\ee

The involutions defined above act on the basis elements as follows:
for any $1\le k\le n$ we have 
\bea
\alpha(\gamma_{a_1\cdots a_k})&=&(-1)^k\gamma_{a_1\cdots a_k}
\,,
\nonumber\\
\tau(\gamma_{a_1\cdots a_k})&=&(-1)^{k(k-1)/2}\gamma_{a_1\cdots a_k}\,,
\nonumber\\
(\gamma_{a_1\cdots a_k})^*&=&(-1)^{k(k+1)/2}\gamma_{a_1\cdots a_k}\,,
\nonumber\\
{\rm Pr}_0(\gamma_{a_1\cdots a_k})&=&0\,.
\eea
In even dimension the chirality operator can be used to define
the main involution $\alpha$
\be
\alpha(\gamma_{a_1\cdots a_k})=\Gamma\gamma_{a_1\cdots a_k}\Gamma
\qquad  {\rm for \ even \ } n\,.
\ee


We list below some properties of the basis elements of the Clifford 
algebra. 
All elements $\gamma_{a_1\dots a_k}$ satisfy the normalization
condition
\be
\tau(\gamma_{a_1\cdots a_k})\gamma_{a_1\cdots a_k}=\II\,,
\ee
and, therefore, are invertible
\be
(\gamma_{a_1\cdots a_k})^{-1}=\tau(\gamma_{a_1\cdots a_k})\,.
\ee
Moreover, the set of elements 
\be
\pm\II, \ \pm\gamma_{a_1},\ \pm\gamma_{a_1a_2}, \ \dots \ , 
\pm\gamma_{a_1\cdots a_{n}},
\qquad
(1\le a_1<a_2<\cdots<a_{n}\le n)
\label{1.196}
\ee
forms a finite multiplicative group.

There holds
\bea
{\rm Pr}_0(\tau(\gamma_{a_1\cdots a_k})\gamma^{b_1\cdots b_j})
&=&0, \qquad {\rm for}\ k\ne j,\\[10pt]
{\rm Pr}_0(\tau(\gamma_{a_1\cdots a_k})\gamma^{b_1\cdots b_k})
&=&k! \delta^{b_1}_{[a_1}\cdots\delta^{b_k}_{a_k]}\,.
\eea
Therefore, there is a natural inner product in the Clifford algebra
defined by
\be
\langle A,B\rangle={\rm Pr}_0 (\tau(A)B)\,.
\ee
The basis introduced above is orthonormal in this inner product.
Thus, every element $A\in \Cliff(n)$ can be presented in the form
\be
A=A_{(0)}\cdot \II
+\sum_{k=1}^{n}{1\over k!}
A_{(k)}^{a_1\cdots a_k}\gamma_{a_1\cdots a_k}\,,
\label{1.198}
\ee
where
\bea
A_{(0)}&=&\langle\II,A\rangle={\rm Pr}_0\, A\,,\\[10pt]
A_{(k)}^{a_1\cdots a_k}&=&
\langle\gamma^{a_1\cdots a_k},A\rangle
={\rm Pr}_0\left(\tau(\gamma^{a_1\cdots a_k})A\right)\,.
\label{at1}
\eea


The product of the basis elements of the Clifford algebra is given by
\cite{zhelnorovich82}
\be
\gamma_{a_1\cdots a_k}\gamma^{b_1\cdots b_j}
=\sum_{p=0}^{n}
(-1)^{p(2k-p-1)/2}{k!j!\over p!(k-p)!(j-p)!}
\delta^{[b_1}_{[a_1}\cdots\delta^{b_p}_{a_p}\gamma_{a_{p+1}\cdots a_k]}
{}^{b_{p+1}\cdots b_j]}
\ee
In particular,
\be
\gamma_{a_1a_2}\gamma^{b_1\cdots b_k}
=\gamma_{a_1a_2}{}^{b_1\cdots b_k}
-2k\delta_{[a_1}^{b_1}\gamma_{a_2]}{}^{b_2\cdots b_k]}
-k(k-1)\delta_{[a_1}^{[b_1}\delta_{a_2]}^{b_2}\gamma^{b_3\cdots b_k]}\,,
\ee
\be
\gamma_{b_1\cdots b_k}\gamma^{a_1a_2}
=\gamma_{b_1\cdots b_k}{}^{a_1a_2}
+2k\delta^{[a_1}_{b_1}\gamma^{a_2]}{}_{b_2\cdots b_k]}
-k(k-1)\delta^{[a_1}_{[b_1}\delta^{a_2]}_{b_2}\gamma_{b_3\cdots b_k]}\,,
\ee
which for $k=2$ takes the form
\be
\gamma_{a_1a_2}\gamma^{b_1b_2}
=\gamma_{a_1a_2}{}^{b_1b_2}
-4\delta_{[a_1}^{b_1}\gamma_{a_2]}{}^{b_2]}
-2\delta_{[a_1}^{[b_1}\delta_{a_2]}^{b_2]}\,.
\ee
Therefore,
\be
[\gamma_{a_1a_2},\gamma^{b_1\cdots b_k}]
=-4k\delta_{[a_1}^{[b_1}\gamma_{a_2]}{}^{b_2\cdots b_k]}\,,
\ee
and, in particular,
\be
[\gamma_{ab},\gamma_{cd}]
=
2\left(
-\delta_{ac}\gamma_{bd}
-\delta_{bd}\gamma_{ac}
+\delta_{bc}\gamma_{ad}
+\delta_{ad}\gamma_{bc}
\right)
\,.
\label{la78}
\ee
Thus $\gamma_{ab}$ form 
a representation of the Lie algebra of the orthogonal group $SO(n)$.

On the other hand, the anti-commutator of the elements $\gamma_{ab}$ is
\be
\gamma_{ab}\gamma_{cd}+\gamma_{cd}\gamma_{ab}
=2\left(\gamma_{abcd}
-\delta_{ac}\delta_{bd}
+\delta_{bc}\delta_{ad}
\right)
\,.
\label{gamma2}
\ee

\subsection{Spin Group}

For any unit vector $u\in V$ we have 
\be
[\gamma(u)]^{-1}={\gamma(u)}\,,
\ee
More generally, let $u_1, \dots u_k$ be a collection of unit vectors 
 from $V$, and let  
\be
T=\gamma(u_1)\cdots\gamma(u_k)\,.
\ee
Then
\bea
\tau(T)T=\II\,.
\eea
Thus, the elements of the Clifford algebra $\Cliff(n)$ of the form
$\gamma(u_1)\cdots\gamma(u_k)$, where $u_1, \dots u_k$ are unit vectors 
from $V$, are invertible and form a multiplicative group
\be
\Pin(n)=\{\II,\gamma(u_1)\cdots\gamma(u_k)\;|\; u_j\in V, 
|u_j|=1, k\in \NN\}
\subset\Cliff(n)\,.
\ee
Alternatively,
\be
\Pin(n)=\{T\in \Cliff(n)\;|\; \tau(T)T=\II, \ 
TC_1(n)T^{-1}\subset C_1(n)\}\,.
\ee

The group $\Pin(n)$ naturally splits into two parts,
\be
\Pin(n)=\Spin(n)\cup {\cal P}\Spin(n)\,,
\ee
an even part $\Spin(n)$, called the spin group, 
consisting of products of even number of elements
\bea
\Spin(n)&=&\{\II,\pm\gamma(u_1)\cdots\gamma(u_{2k})\;|\; u_j\in V, 
|u_j|=1, k\in \NN\}
\subset\Cliff_+(n)
\nonumber\\[10pt]
&=&\{T\in \Cliff_+(n)\;|\; \tau(T)T=\II, \ 
TC_1(n)T^{-1}\subset C_1(n)\}\,,
\eea
and the odd part  ${\cal P}\Spin(n)$
consisting of products of odd number of elements
$\gamma(u_1)\cdots\gamma(u_{2k+1})$ (which do not form a group). 
Here ${\cal P}=\gamma(e)$ with some unit vector $e$.
It is easy to see that the group $\Pin(n)$ is generated 
multiplicatively by reflections in all hyperplanes.
The spin group $\Spin(n)$ is the subgroup of the group $\Pin(n)$
generated by even number of reflections. 

Let $u\in V$ and $T\in \Pin(n)$.
Then there is a vector $v\in V$ such that
\be
T\gamma(u) T^{-1}=\gamma(v)\,.
\ee
This defines an orthogonal transformation of $V$
\be
\tilde\rho(T): u\mapsto v=\tilde\rho(T)u\,.
\ee
We slightly modify this definition by including an additional factor
\be
\tilde \rho(T)=(\alpha\circ\rho)(T)=\varepsilon(T)\rho(T)\,,
\ee
where $\alpha$ is the main involution and 
$\varepsilon(T)$ is the parity of the element $T$. Hence,
this modification does not affect the spin group.
Clifford algebra $\Cliff(n)$ carries a natural action of the orthogonal
group $O(n)$ inherited from the tensor algebra.
Thus the homomorphism $\rho$ is defined by
\be
\varepsilon(T)T\gamma(u)T^{-1}=\rho(T)\gamma(u)\,,
\ee
or
\be
T\gamma(u)T^*=\rho(T)\gamma(u)\,.
\ee
In particular,
\be
\varepsilon(T)T\gamma^a T^{-1}=\rho^a{}_b(T)\gamma^b\,.
\ee
Hence there is a continuous surjective two-to-one homomorphism
\be
\rho: \Pin(n)\to O(n)\,, 
\ee
defined by
\be
\rho^a{}_b(T)=\varepsilon(T){\rm Pr}_0(T\gamma^aT^{-1}\gamma_b)
={\rm Pr}_0(T\gamma^a T^*\gamma_b)
\ee
so that 
\be
O(n)=\Pin(n)/\ZZ_2\,.
\ee
Similarly,
\be
\rho: \Spin(n)\to SO(n)\,,
\ee
is a continuous surjective two-to-one homomorphism and
\be
SO(n)=\Spin(n)/\ZZ_2\,.
\ee

This means that the group $\Pin(n)$ is a double covering group of the 
orthogonal group $O(n)$. The group $O(n)$ is disconnected and has two
connected components: the proper subgroup  $SO(n)$ containing the proper
orthogonal transformations (with determinant equal to $+1$),  and
$PSO(n)$ consisting of orthogonal transformations with determinant
equal to $(-1)$. The elements of $PSO(n)$ are products of a proper
orthogonal transformation from $SO(n)$ and a reflection $P$.  Thus,
\be
O(n)=SO(n)\cup PSO(n)\,.
\ee
The group $SO(n)$ is connected but not simply connected.  The spin group
$\Spin(n)$ is a double covering group of the special orthogonal group
$SO(n)$. For $n=2$ the group $\Spin(2)$ is connected but not  simply
connected, whereas for $n\ge 3$ the group $\Spin(n)$ is simply 
connected and is the universal covering group of $SO(n)$.

The eq. (\ref{la78}) shows that the space $C_2(n)$ is closed under the
algebra commutator. Therefore, it forms a Lie algebra with the Lie
bracket identified with the Clifford algebra commutator. This Lie
algebra is the Lie algebra of the spin group $\Spin(n)$. The Lie algebra
of the group $\Pin(n)$ is, of course, the same. The generators of this
Lie algebra are the basis elements $\gamma_{ab}$, which 
form a representation of the Lie algebra of the orthogonal
group $SO(n)$. Thus, the Lie algebra of the spin group is isomorphic to the
Lie algebra of the orthogonal group $SO(n)$.  

In other words, the spin group ${\rm
Spin}(n)$ is obtained by exponentiating the
Lie algebra of the group $SO(n)$ inside the Clifford algebra 
\be
\Spin(n)=\exp[C_2(n)]\,.
\ee
Let $\theta$ be an element of the Lie algebra of the group  $SO(n)$
represented by an antisymmetric matrix $(\theta_{ab})$. 
Then $\theta_{ab}\gamma^{ab}\in C_2(n)$ is an
element of the Lie algebra of the spin group $\Spin(n)$ and the double
covering homomorphism $\rho: \Spin(n)\to SO(n)$ is given by
\be
\rho\left[\exp\left(-{1\over 4}\theta_{ab}\gamma^{ab}\right)\right]
=\exp(\theta)\,.
\label{sp}
\ee

\subsection{Spin Representation}

Recall that $N=2^m$.
Let $S$ be a $N$-dimensional complex vector space (which is, of
course, isomorphic to $\CC^{N}$),  $S^*$ be the dual space of linear
functionals $S\to \CC$ and $\End(S)$ be the  algebra of linear 
endomorphisms $S\to S$ of the vector space $S$ (which is isomorphic to
the  vector space $\Mat(N,\CC)$ of complex square matrices
of order $N$).
We will call the elements of the vector space $S$ the Dirac  spinors (or
complex spinors). Let $\left<\ , \ \right>: S\times S\to \CC$  be an 
inner product on $S$. Then the elements of the dual space $S^*$ are
naturally identified with the adjoint vectors by
\be
\bar\psi(\varphi)=\left<\psi,\varphi\right>\,,
\ee
the space of endomorphisms $\End(S)$ is identified with $S\otimes S^*$,
and the adjoint $\bar T$ of an endomorphism $T$ is defined with respect
to this inner product, that is
\be
\left<\psi, T\varphi\right>=\left<\bar T\psi,\varphi\right>\,.
\ee
Finally, we denote by $\Aut(S)$ the group of automorphisms (invertible
linear 
endomorphisms) of the  vector space $S$ (which is isomorphic to the 
general linear group $GL(N,\CC)$ of complex non-degenerate
square matrices of order $N$) 
and by $U(S)$ the group of unitary endomorphisms
(which is isomorphic to $U(N)$)  that preserve the inner product,
that is
\be
\bar UU=\II\,.
\ee

Then, in even dimension $n=2m$ 
the complex Clifford algebra $\Cliff_\CC(2m)$ is isomorphic to the
algebra of endormorphisms $\End(S)$
\be
\Cliff_\CC(2m)=\End(S)\,.
\ee
In odd dimension $n=2m+1$ the Clifford algebra $\Cliff_\CC(2m+1)$
is isomorphic to the direct sum of two copies of $\End(S)$
\be
\Cliff_\CC(2m+1)=\End(S)\oplus\End(S)\,.
\ee

Thus in even dimension $n=2m$ one can  identify the elements of the
complex Clifford algebra $\Cliff_\CC(2m)$ with the complex square
matrices of order $N$. The unit element is identified with the 
unit matrix and the elements $\gamma_a$ become then the
Dirac matrices. 

In odd dimension $n=2m+1$ the dimensionality of the representation space
should be doubled.  That is the elements of the complex  Clifford
algebra $\Cliff_\CC(2m+1)$ if odd dimension $n=2m+1$ are identified with
the  complex block matrices of order $2N$. Of course,  now the
unit element is the unit matrix of order $2N$. Let
$\{\gamma'_a\}$, where $a=1,\dots,2m,$ be the Dirac matrices of
order $N$ in even dimension $n=2m$ and $\Gamma'$ be the
corresponding chirality operator.  Then the elements $\gamma_a$ of the
complex  Clifford algebra $\Cliff_\CC(2m+1)$ if odd dimension $n=2m+1$
are 
\be
\gamma_{a}=\left(
\begin{array}{cc}
\gamma'_a & 0\\
0 & \gamma'_a
\end{array}
\right)\,, \qquad
\gamma_{2m+1}=\left(
\begin{array}{cc}
\Gamma' & 0\\
0 & -\Gamma'
\end{array}
\right)\,.
\ee
The basis elements in odd dimension are
\be
\gamma_{a_1\cdots a_k}=\left(
\begin{array}{cc}
\gamma'_{a_1\cdots a_k} & 0\\
0 & \gamma'_{a_1\cdots a_k}
\end{array}
\right)\,,
\ 
\gamma_{a_1\cdots a_k, (2m+1)}=\left(
\begin{array}{cc}
\gamma'_{a_1\cdots a_k}\Gamma' & 0\\
0 & -\gamma'_{a_1\cdots a_k}\Gamma'
\end{array}
\right)\,,
\ee
where $1\le k\le 2m$ and the indices $a_j$ run over $1,\dots,2m$. The
unit matrix and the chirality operator in odd dimension, which determine
the center of the Clifford algebra $\Cliff_\CC(2m+1)$ in odd dimension,
are
\be
\II=\left(
\begin{array}{cc}
\II' & 0\\
0 & \II'
\end{array}
\right)\,,
\qquad
\Gamma=\left(
\begin{array}{cc}
\II' & 0\\
0 & -\II'
\end{array}
\right)\,.
\ee

Note that the projection ${\rm Pr}_0$ onto the identity element is
nothing but the matrix trace normazed so that ${\rm Pr}_0(\II)=1$.

The spin representation of the Clifford algebra $\Cliff(n)$ is a 
representation with the representation space $S$, that is with complex
square 
matrices of order $N$. Thus in even dimension $n=2m$ there is only
one irreducible faithful representation of the Clifford algebra
$\Cliff(2m)$. In odd dimension $n=2m+1$  the spin representation of the
Clifford algebra $\Cliff_\CC(2m+1)$ is obtained by an additional
projection onto either the first or the second component of
$\End(S)\oplus\End(S)$. Thus there are two non-equivalent faithful
irreducible spin representations of the Clifford algebra
$\Cliff_\CC(2m+1)$ by complex square 
matrices of order $N$, one obtained by
the set of matrices $\{\gamma'_a,\Gamma'\}$ and the other by the set of
matrices  $\{\gamma'_a,-\Gamma'\}$.

Since the spin group $\Spin(n)$ is embedded in the Clifford algebra
$\Cliff_\CC(n)$, this also defines the spin representation of the spin
group. The elements of the Clifford algebra act on the vector space $S$,
and, therefore, $S$ becomes the Clifford module, that is a module  over
the Clifford algebra.

In even dimension the chirality operator $\Gamma$ is a nontrivial
involution, which has the eigenvalues $+1$ and $-1$. Since  it is not in
the center of the Clifford algebra, it splits the whole spinor space $S$
into the eigenspaces $S_+$ and $S_-$ corresponding to  these
eigenvalues. Thus the spin representation of the spin group
$\Spin(2m)$ in even dimension decomposes
into the eigenspaces of the chirality operator, that is
\be
S=S_+\oplus S_-\,,
\ee
where the subspaces $S_{\pm}$ are defined by
\be
S_{\pm}=\{\psi\in S\;|\;\Gamma\psi=\pm\psi\}\,.
\ee
The spinors from the spaces $S_+$ and $S_-$ are called
right and left (or positive and negative) Weyl spinors
(or half-spinors) respectively.
Of course, the dimension of the subspaces $S_\pm$ is equal to one half
of the dimension of the space $S$
\be
\dim S_\pm={N\over 2}\,.
\ee 
Also, since $\Gamma$ anticommutes with $\gamma_a$, 
the Clifford multiplication intertwines the chiral subspaces
\be
C_1(n)S_\pm=S_\mp\,.
\ee

In odd dimension the chirality operator is trivial, it is either
$\Gamma=+\II$ or $\Gamma=-\II$, depending on the spin representation
of the Clifford algebra. Therefore,  there is only one irreducible
spin representation of the spin group $\Spin(2m+1)$ in odd dimension,
i.e. there are no half-spinors in odd dimension.

Finally, in the spinor space $S$  there exists a Hermitian
positive-definite inner product  $\left<\;,\;\right>$ such that for any
unit vector $u\in V$ the element $\gamma(u)$ is self-adjoint and
unitary
\be
\bar\gamma(u)=\gamma(u)=[\gamma(u)]^{-1}\,.
\ee
In this representation the chirality operator $\Gamma$ is also self-adjoint
and unitary
\be
\bar \Gamma=\Gamma=\Gamma^{-1}\,.
\ee
In even dimension, the chiral subspaces $S_+$ and $S_-$ are orthogonal
in this inner product.


\subsection{Spin Connection}

Let $(M,g)$ be a smooth compact orientable $n$-dimensional Riemannian
spin manifold without boundary and with a  positive-definite Riemannian
metric $g$.   Let the tangent bundle $TM$ be oriented by choosing a
smooth oriented basis. Since $M$ is orientable the transtion functions
are matrices from $SO(n)$. Let $SO(M)$ be the frame bundle,  i.e. the
principal fiber bundle of oriented orthonormal frames  with the
structure group $SO(n)$. The typical fiber of the frame bundle $SO(M)$
is $SO(n)$. The spin group $\Spin(n)$ is a double cover of the group
$SO(n)$ (for $n\ge 3$ it is the universal cover and, thus, simply
connected). A spin structure on $M$ is a principal bundle $\Spin(M)$
with the structure group $\Spin(n)$ together with a double covering
homomorphism ${\rm Spin}(M)\to SO(M)$ which preserves the group action.
The necessary and sufficient conditions for a manifold to have a spin
structure are the vanishing of the first two Stiefel-Whitney classes of
the manifold $M$. There can be several possible inequivalent spin
structures, which are parametrized by representations of the fundamental
group $\pi_1(M)$. For simply connected manifolds the spin structure is
unique. 

The spinor bundle ${\cal S}$ is the associated vector bundle
with the structure group $\Spin(n)$ whose typical fiber is the spinor
space $S$. Spinor fields are  sections of this vector bundle.  We denote
by $C^{\infty}({\cal S})$ the space of smooth sections of the spinor
bundle. Using the Hermitian inner product $\left<\;,\;\right>$ on the
spinor space $S$ and  the invariant Riemannian volume element $d\,\vol$
on $M$  one defines the natural $L^2$-inner product $(\;,\;)$ in
$C^\infty({\cal S})$ and the Hilbert space of square integrable sections
$L^2({\cal S})$ as the completion of $C^\infty({\cal S})$ in this norm.


To define the Dirac operator on a Riemannian manifold $M$ we need a
connection (covariant derivative) on the spinor bundle ${\cal S}$
\be
\label{nv} 
\nabla^{\cal S}: \
C^\infty({\cal S})\to C^\infty(T^*M\otimes {\cal S})\,,
\ee
which we assume to be compatible with the Hermitian inner product on the 
spinor bundle ${\cal S}$. This connection is naturally extended to
bundles in the tensor algebra over ${\cal S}$ and ${\cal S}^*$. Any
Riemannian manifold has a unique symmetric connection $\nabla^{TM}$
compatible with the metric, the Levi-Civita connection.  In fact, using
the Levi-Civita connection together with $\nabla^{\cal S}$, we naturally
obtain  connections on bundles in the tensor algebra over ${\cal
S},\,{\cal S}^*,\,TM,\,T^*M$; the resulting connection will be
denoted just by $\nabla$. It will usually be clear which bundle's
connection is being referred to, from the nature of the section being
acted upon. 


All the homomorphism and involutions of the Clifford algebra are naturally
extended to bundle maps, in particular,
\be
\gamma: T^*M\to {\cal S}\,.
\ee
Since the
pricipal bundle $\Spin(M)$ is a double cover of the orthonormal frame
bundle $SO(M)$, it inherits the Levi-Civita connection. The exact form
of this correspondence is obtained from the differential of the
homomorphism (\ref{sp}) 
\be
\rho: \Spin(M)\to SO(M)\,.
\ee
Since for any two
spinors $\psi,\varphi\in C^\infty({\cal S})$,
 $\langle\psi,\gamma_{a_1\cdots a_k}\varphi\rangle
\in C^\infty(\wedge^k T^*M)$
transforms like a tensor (in fact, like a $k$-form),
then the spin connection can be defined by
requiring it to satisfy the Leibnitz rule
\be
\nabla_b\langle\psi,\gamma_{a_1\cdots a_k}\varphi\rangle
=\langle\nabla_b\psi,\gamma_{a_1\cdots a_k}\varphi\rangle
+\langle\psi,\gamma_{a_1\cdots a_k}\nabla_b\varphi\rangle\,.
\ee

We label the local coordinates $x^\mu$ on the manifold $M$ by Greek
indices which run over $1,\dots, n$.  Let $\partial_\mu$ be a coordinate
basis for the tangent space $T_xM$ at a point $x\in M$  and let
$\gamma_\mu=\gamma(\partial_\mu)$. Then 
\be
\gamma_\mu\gamma_\nu+\gamma_\nu\gamma_\mu=2g_{\mu\nu}\II\,,
\ee
where $g_{\mu\nu}=g(\partial_\mu,\partial_\nu)$ 
is the Riemannian metric.
Let 
\be
e_a=e^\mu_a\partial_\mu
\ee
be an orthonormal basis for the tangent space
$T_xM$. 
Let $e^a_\mu$ be the matrix inverse to $e^\mu_a$, defining the dual
basis
\be
\omega^a=e^a_\mu dx^\mu\,
\ee
in the cotangent space $T_x^*M$.
Then  
\be
g_{\mu\nu}e^\mu_a e^\nu_b=\delta_{ab}\,,
\qquad
g^{\mu\nu}e_\mu^a e_\nu^b=\delta^{ab}\,,
\ee
and the matrices $\gamma_\mu$ are related to the constant
matrices $\gamma_a$ forming a representation of the Clifford algebra by
\be
\gamma_a=e^\mu_a\gamma_\mu\,,
\qquad
\gamma_\mu=e^a_\mu\gamma_a\,.
\ee
Similarly, we define
\be
\gamma_{\mu_1\cdots\mu_k}=\gamma_{a_1\cdots a_k}e^{a_1}_{\mu_1}
\cdots e^{a_k}_{\mu_k}\,.
\ee
Thus, in local coordinates one obtains for the spin connection
\be
\nabla_\mu\psi
=\left(\partial_\mu+{1\over 4}\gamma_{ab}\omega^{ab}{}_{\mu}\right)\psi\,,
\ee
where $\omega^{ab}{}_{\mu}$ is the spin connection one-form defined by
\be
\omega^{ab}{}_{\mu}=
e^{a\nu}\partial_{[\nu} e_{\mu]}^b
-e^{b\nu}\partial_{[\nu} e_{\mu]}^a
+e_{c\mu}e^{a\nu}e^{b\lambda}\partial_{[\nu} e_{\lambda]}^c\,.
\ee
This is nothing but the Fock-Ivanenko coefficients 
\cite{fockivanenko29a, fockivanenko29b}.

We will generalize the above setup as follows.
Let $G$ be a compact semi-simple Lie group and ${\cal
G}$ be the principal fiber bundle over the manifold $M$  with the
structure group $G$. Let  ${\cal W}$ be the associated vector bundle
with the structure group $G$ whose typical fiber is a vector space $W$. 
Then the vector bundle ${\cal W}\otimes {\cal S}$ is a twisted
spinor bundle. The sections of the twisted spinor bundle are represented
locally by $k$-tuples of spinors, where $k=\dim W$ is the dimension of the
vector space $W$. 
For a twisted spinor bundle ${\cal W}\otimes {\cal S}$ 
the covariant derivative is defined by
\be
\nabla_\mu\psi
=\left(\partial_\mu+{1\over 4}\gamma_{ab}\omega^{ab}{}_{\mu}
+{\cal A}_\mu\right)\psi\,,
\ee
where ${\cal A}_\mu$ is the connection $1$-form on the vector bundle 
${\cal W}$ taking values in the Lie algebra of the gauge group $G$. In
the following, we redefine the definition of the spinor bundle. We will
denote the twisted spinor bundle ${\cal W}\otimes {\cal S}$ by ${\cal
S}$ and call it just the spinor bundle. The meaning of the bundle
(twisted or not) is usually clear from the context. Note that the
dimension of the fiber of the twisted spinor bundle is 
$2^m\cdot\dim W\,$.
So, when dealing with the twisted spinor bundle we will redefine the
definition of the number $N$. It will asways mean the dimension of the
fiber of the spinor bundle, whether twisted or not.

The curvature of the spin connection is described by 
the commutator of the covariant derivatives
\be
[\nabla_\mu,\nabla_\nu]\psi={\cal R}_{\mu\nu}\psi\,,
\ee
where
\be
{\cal R}_{\mu\nu}
={1\over 4}\gamma^{\alpha\beta}R_{\alpha\beta\mu\nu}
+{\cal F}_{\mu\nu}\,,
\label{124}
\ee
$R_{\alpha\beta\mu\nu}$ is the Riemann curvature of the metric $g$ and
\be
{\cal F}_{\mu\nu}=\partial_\mu A_\nu-\partial_\nu{\cal A}_\mu
+[{\cal A}_\mu,{\cal A}_\nu]\,.
\ee

\subsection{Dirac Operator}

The Dirac operator is a first order partial differential operator
acting on smooth sections of the spinor bundle
\be
D: C^\infty({\cal S})\to C^\infty({\cal S})
\ee
defined by the composition of the covariant
derivative with the Clifford multiplication
\be
D=i\gamma\nabla=i\gamma^\mu\nabla_\mu\,.
\ee
The leading symbol of the Dirac operator is
\be
\sigma_L(D;x,\xi)=-\gamma^\mu(x)\xi_\mu\,,
\ee
where $\xi\in T^*_xM$ is a covector at a point $x\in M$.  Since it is
self-adjoint and  non-degenerate for any $\xi\ne 0$, $x\in M$, the Dirac
operator $D$ is elliptic.  One can also easily check that the Dirac
operator is symmetric,  (or formally self-adjoint), 
that is, for any two smooth spinor fields
$\psi,\varphi\in C^\infty({\cal S})$
\be
(D\psi,\varphi)=(\psi,D\varphi)\,.
\ee

The Laplacian is a second order partial differential operator
acting on smooth sections of the spinor bundle
\be
\Delta: C^\infty({\cal S})\to C^\infty({\cal S})
\ee
defined by
\be
\Delta=-\overline{\nabla}\nabla=g^{\mu\nu}\nabla_\mu\nabla_\nu\,,
\ee
where
\be
\overline{\nabla}: C^\infty(T^*M\otimes {\cal S})\to C^\infty({\cal S}) 
\ee
is the formal adjoint of the covariant derivative operator
with respect to the $L^2$ inner product on the spinor bundle 
${\cal S}$.

The square of the Dirac operator is 
\be
D^2=-\Delta-{1\over 2}\gamma^{\mu\nu}{\cal R}_{\mu\nu}\,.
\ee
By using the curvature of the spin connection (\ref{124}),
the eq. (\ref{gamma2}) and the Bianci identity
we obtain the Lichnerowicz formula
\be
D^2=-\Delta+{1\over 4}R\,\II-{1\over 2}\gamma^{\mu\nu}{\cal F}_{\mu\nu}\,,
\ee
where $R$ is the scalar curvature.

The leading symbol of the operator $D^2$ 
\be
\sigma_L(D^2;x,\xi)=g^{\mu\nu}(x)\xi_\mu\xi_\nu\,\II\,
\ee
is, of course, elliptic, self-adjoint, scalar and positive-definite. 

The Dirac operator $D$ is a formally  self-adjoint elliptic operator 
acting on smooth sections of spinor bundle over a compact manifold
without boundary. One can show that $D$  is essentially self-adjoint,
that is, its closure is self-adjoint and, hence, it has a unique
self-adjoint extension to $L^2({\cal S})$. The same is true for its
square $D^2$. It is well known that the operator $D$ has a discrete
real spectrum $(\lambda_n)_{n=1}^\infty$, which can be ordered
according to
\be
0\le \lambda^2_1\le\lambda^2_2\le\cdots\le\cdots\le\lambda^2_n\le\cdots\,.
\ee
Moreover, each eigenspace is finite-dimensional and the eigenspinors
$(\varphi_n)_{n=1}^\infty\in C^\infty({\cal S})$ 
are smooth sections of the spinor bundle
that form an orthonormal basis in $L^2({\cal S})$.

Let the dimension of the manifold $n=2m$ be even. Then the spinor bundle
${\cal S}$ has a $\ZZ_2$ grading 
\be
{\cal S}={\cal S}_+\oplus{\cal S}_-\,,
\ee
where ${\cal S}_\pm$ are the subbundles of the right (left) Weyl spinors.
It is easy to see that the Dirac operator anticommutes with the chirality 
operator 
\be
\Gamma D=-D\Gamma
\ee
and, therefore, interchanges the parity of the spinors, that is, in fact,
\be
D: C^\infty({\cal S}_{\pm})\to C^\infty({\cal S}_{\mp})\,.
\ee
In other words, the Dirac operator has odd parity, and, therefore, 
its square $D^2$ is an even operator
\be
D^2: C^\infty({\cal S}_{\pm})\to C^\infty({\cal S}_{\pm})\,.
\ee
Let
\be
P_{\pm}={1\over 2}(\II\pm\Gamma)
\ee
be the projections onto the subbundles ${\cal S}_\pm$
and
\be
D_\pm=P_\mp DP_\pm\,.
\ee
Then
\be
D_\pm^2=0\,, \qquad
D_+=\overline{D_-}\,.
\ee
and 
\be
D=D_++D_-, \qquad 
D^2=D_-D_++D_+D_-\,.
\ee
The operators
\be
D_\mp D_\pm: C^\infty({\cal S}_\pm)\to C^\infty({\cal S}_\pm)\,
\ee
are second-order self-adjoint non-negative
differential operators of even parity
\be
\overline{D_\mp D_\pm}=D_\mp D_\pm\,.
\ee

Thus, $D^2$ acts in the chiral subbundles of the spinor bundle ${\cal
S}_\pm$. One can easily show that for all non-zero eigenvalues there is
an isomorphism between the right and left eigenspaces. In particular,
their dimensions, that is the multiplicities $d_n^{\pm}$ of the right
and left eigenspinors corresponding to the same  non-zero eigenvalue
$\lambda_n^2$, are equal. This clearly does not work for the zero
eigenvalues; so there could be any number of right or left eigenspinors
corresponding to zero eigenvalue.

Let 
\be
\Ker(D)=\{\psi\in C^\infty({\cal S})\;|\;D\psi=0\}
\ee
be the kernel of the operator $D$,
that is the vector space of its zero eigenspinors. 
Then
\be
\Ker(D_{\pm})=\Ker(D)\cap C^\infty({\cal S}_{\pm})
=\{\psi\in C^\infty({\cal S}_\pm)\;|\;D\psi=0\}
\ee
are invariant subspaces of the right and left zero eigenspinors
of the Dirac operator and
\be
\Ker(D)=\Ker(D_+)\oplus\Ker(D_-)\,.
\ee
The index of the 
Dirac operator is a topological invariant of the manifold $M$
and the spinor bundle defined by
\be
{\rm Ind}(D)=\dim \Ker(D_+)-\dim\Ker(D_-)\,.
\ee

\subsection{Heat Kernel}

Thus, $D^2$ is a self-adjoint elliptic second-order
partial differential operator with a
positive definite scalar leading 
symbol acting on sections of spinor bundle over
a compact manifold without boundary. Such operators are called Laplace
type operators. 
For $t>0$ the heat semigroup 
\be
\exp(-tD^2): L^2({\cal S})\to L^2({\cal S})
\ee
is a bounded operator (in fact, it is 
a smoothing operator $L^2({\cal S})\to C^\infty({\cal S})$). 
The integral kernel of this operator, called the heat kernel, is
\be
U(t;x,x')=\sum_{n=1}^\infty e^{-t\lambda_n^2}\varphi_n
\otimes\bar\varphi_n(x')\,,
\ee
where each eigenvalue is counted with its multiplicity.
The heat kernel satisfies the heat equation
\be
(\partial_t+D^2)U(t;x,x')=0
\ee
with the initial condition
\be
U(0^+;x,x')=\delta(x,x')\,,
\ee
where $\delta(x,x')$ is the Dirac distribution.

For $t>0$ the heat kernel $U(t;x,x')$ is a smooth function near the
diagonal of $M\times M$ and has a well defined diagonal value
$U(t;x,x)$.
Moreover, the heat semigroup is a trace-class operator with a well
defined $L^2$-trace
\be
\Tr_{L^2}\exp(-tD^2)=\int_M d\vol(x) \tr_{S}U(t;x,x)\,,
\ee
where $\tr_{S}$ is the trace in the spinor space.
The trace of the heat kernel is a spectral invariant of the Dirac operator
since
\be
\Tr_{L^2}\exp(-tD^2)=\sum_{n=1}^\infty e^{-t\lambda_n^2}\,.
\ee

Similarly, let $F\in C^\infty(\End({\cal S}))$ be a smooth section
of the endomorphism bundle of the spinor bundle. We can define
the trace
\be
\Tr_{L^2}\left[F\exp(-tD^2)\right]=\int_M d\vol(x) 
\tr_{S}\left[F(x)U(t;x,x)\right]\,.
\ee
Note, however, that, in general, this is not a spectral invariant.

In a particular case, when the dimension of the manifold is even and 
$F$ is the chirality operator, $F=\Gamma$, we obtain
\be
\Tr_{L^2}\left[\Gamma\exp(-tD^2)\right]
=\Tr_{L^2}\exp(-tD_-D_+)-\Tr_{L^2}\exp(-tD_+D_-)
\,.
\ee
Since the nonzero spectra of the operators $D_-D_+$ and $D_+D_-$ are
isomorphic, we obtain
\be
\Tr_{L^2}\left(\Gamma\exp(-tD^2)\right)={\rm Ind}(D)\,.
\ee
That is, this trace does not depend
on $t$ and is a topological invariant equal to the index of 
the operator $D$.

One can show that there is an asymptotic expansion of the diagonal
of the heat kernel as $t\to 0$~\cite{gilkey95}
(for a review, see also 
\cite{avramidi91b,avramidi99,avramidi00,avramidi02})
\begin{equation}
U(t;x,x)\sim (4\pi)^{-n/2} \sum_{k=0}^\infty t^{(2k-n)/2}
a_k(D^2;x)\,,
\end{equation}
and the corresponding expansion of the trace of the
heat kernel
\be
\Tr_{L^2}\left[F\exp(-tD^2)\right]
\sim (4\pi)^{-n/2}\sum_{k=0}^\infty t^{(2k-n)/2} A_k(F,D^2)\,.
\ee
The coefficients $A_{k}(F,D^2)$, called the global heat invariants,
are invariants determined by the integrals over the manifold
\be
A_k(F,D^2)=\int_M d\vol(x)\;\tr_{S}F a_k(D^2;x)\,,
\ee
of local heat invariants $a_k(D^2;x)$ constructed polynomially 
from the jets of the
symbol  of the Dirac operator $D$, so that they are polynomial in
curvatures and their covariant derivates.

In the particular case $F=\II$ the heat invariants $A_k(\II,D^2)$ are
spectral invariants of the Dirac operator, and in the case $F=\Gamma$
(in even dimension $n$) all heat invariants $A_k(\Gamma,D^2)$
vanish except for one that determines the index of the Dirac operator,
that is
\bea
A_k(\Gamma,D^2)&=&0, \qquad\qquad\qquad\qquad k\ne {n\over 2}\,,
\\[10pt]
A_{n/2}(\Gamma, D^2)&=&(4\pi)^{n/2}{\rm Ind}(D)\,.
\eea

The first two spectral invariants are given by
\bea
A_0(\II,D^2)&=&N\int_M d\vol\;1\,,
\label{170}
\\[10pt] 
A_1(\II,D^2)&=&-{1\over 12}N\int_M d\vol\; R\,.
\label{171}
\eea
where $N=\dim S$.


\section{Dirac Operators in Matrix Geometry}

In this section we closely follow our papers 
\cite{avramidi03,avramidi04a,avramidi04b}.

\subsection{Non-commutative Metric and Dirac Matrices}

Now, let $S$ be a $N$-dimensional complex vector space with a positive
definite Hermitean inner product $\langle \;,\,\rangle$, $S^*$ be its
dual vector space and $\End(S)$ be the space of linear  endomorphisms of
the vector space $S$. The vector space $S$ is isomorphic to $\CC^N$ and
$\End(S)$ be  is isomorphic to the vector space $\Mat(N,\CC)$ of complex
square matrices of order $N$.  The group of automorphisms $\Aut(S)$ of the
vector space $S$ is isomorphic to the general linear group $GL(N,\CC)$
of complex square 
nondegenerate matrices of order $N$
and the group of unitary
endomorphisms $G(S)$ is isomorphic to $SU(N)$; the dimension of the group
$G$ is $\dim G=N^2$.
The group $G$ acts on vectors and covectors by left and right action 
\bea
\varphi'&=&U\varphi\,,\qquad \varphi \in S
\\
\bar\varphi'&=&\bar\varphi U^{-1}\,, \qquad \bar\varphi \in S^*\,.
\label{gt0}
\eea

Now, let $M$ be a smooth compact orientable $n$-dimensional manifold
without boundary and 
\be
V=T_xM \qquad
{\rm and}\qquad
V^*=T_x^*M
\ee
be the tangent and
contangent spaces at a point $x$ in $M$. We introduce the following
notation for the vector spaces of vector-valued and endomorphism-valued
tensors
\be
\Lambda_p=\wedge^p V^*\otimes S\,,\qquad
\Lambda^p=\wedge^p V\otimes S\,.
\ee
\be
E_p=\wedge^p V^*\otimes \End(S)\,,\qquad
E^p=\wedge^p V\otimes \End(S)\,.
\ee
Suppose we are given a map
\be
\Gamma: V^*\to \End(S)
\ee
determined by a self-adjoint endomorphism-valued vector
$\Gamma\in V\otimes\End(S)$ given locally by the
matrix-valued vector $\Gamma^\mu$.
Let us define an endomorphism-valued tensor
$a\in V\otimes V\otimes \End(S)$ by
\begin{equation}
a^{\mu\nu}={1\over 2}\left(\Gamma^\mu\Gamma^\nu
+\Gamma^\nu\Gamma^\mu\right)\,.
\end{equation}
Then $a^{\mu\nu}$ is self-adjoint and symmetric
\be
a^{\mu\nu}=a^{\nu\mu}\,\qquad
\overline{a^{\mu\nu}}=a^{\mu\nu}\,.
\ee

One of our main assumptions about the matrix $a$ is that it defines an
isomorphism between the 
spaces $\Lambda_1=V^*\otimes S$ and $\Lambda^1=V\otimes S$, i.e.\
\begin{equation}
a: \Lambda_1\to \Lambda^1\,.
\label{16}
\end{equation}
Let us consider the matrix
\begin{equation}
H(\xi)=a^{\mu\nu}\xi_\mu\xi_\nu=[\Gamma(\xi)]^2\,,
\label{24}
\end{equation}
with $\xi\in T_x^*M$ being a cotangent vector and
$\Gamma(\xi)=\Gamma^\mu\xi_\mu$. Our second assumption is
that this matrix is positive definite, i.e.
\be
H(\xi)>0 \quad {\rm for \ any \ } \xi\ne 0\,.
\ee
Thus, all eigenvalues of this matrix are real and positive for $\xi\ne 0$.
We will call the matrix $a^{\mu\nu}$ the non-commutative metric
and the matrices $\Gamma^\mu$ the non-commutative Dirac matrices.

We will also need a self-adjoint non-degenerate endomorphism
$\rho\in \End(S)$ (given locally by a matrix-valued function).
In the case when $S$ is a spinor space described in section 2 there is
a very simple particular solution
\begin{equation}
\Gamma^\mu=\gamma^\mu\,,\qquad
a^{\mu\nu}=g^{\mu\nu}\II\,, \qquad \rho=g^{1/4}\II\,,
\end{equation}
where $\gamma^\mu$ are Dirac matrices in a Riemannian manifold with a
Riemannian metric $g^{\mu\nu}$ and
\begin{equation}
g=|\det g^{\mu\nu}|^{-1}\,.
\end{equation}
These matrices satisfy all the above conditions.  
We will refer to this particular
case as the \emph{commutative limit}.
In general, we represent these objects as a {\it deformation} of the
commutative limit
\be
\Gamma^\mu=\gamma^\mu+\varkappa\alpha^\mu\,,\qquad
a^{\mu\nu} = g^{\mu\nu}\II+\varkappa h^{\mu\nu}\,,\qquad
\rho = g^{1/4} \exp(\varkappa\phi)\,,
\ee
where $\varkappa$ is a deformation papameter, $\alpha^\mu$
and $\phi$ are some matrices and
and
\be
h^{\mu\nu}=(\alpha^\mu\gamma^\nu+\gamma^\nu\alpha^\mu)
+(\alpha^\nu\gamma^\mu+\gamma^\mu\alpha^\nu)
+\varkappa(\alpha^\mu\alpha^\nu+\alpha^\nu\alpha^\mu)\,.
\ee
Our construction should make
sense in the limit $\varkappa\to 0$ as a power series in the deformation 
papameter.


Since the map $a$~(\ref{16}) is an isomorphism, the inverse map
\begin{equation}
b=a^{-1}: \Lambda^1\to \Lambda_1\,,
\end{equation}
is well defined. In other words, for any $\psi\in \Lambda^1$
there is a unique $\varphi_\nu\in\Lambda_1$ satisfying the
equation $a^{\mu\nu}\varphi_\nu=\psi^\mu$, and, therefore, there is a
unique solution of the equations
\begin{equation}
a^{\mu\nu}b_{\nu\alpha}=\delta^\mu_{\alpha}\,,\qquad
b_{\alpha\nu}a^{\nu\mu}=\delta^\mu_{\alpha}\,.
\end{equation}
Notice that the matrix $b_{\mu\nu}$ has the property
\begin{equation}
\bar b_{\mu\nu} = b_{\nu\mu}\,,
\end{equation}
but is neither symmetric $b_{\mu\nu}\ne b_{\nu\mu}$ nor self-adjoint
$\bar b_{\mu\nu}\ne b_{\mu\nu}$.

The isomorphism $a$ naturally defines the maps
\begin{equation}
A: \Lambda_p\to \Lambda^{p}\,,\qquad 
B: \Lambda^p\to \Lambda_{p}\,,
\end{equation}
as follows
\begin{equation}
(A\varphi)^{\mu_1\cdots\mu_p} =A^{\mu_1\cdots\mu_p\nu_1\cdots\nu_p}
\varphi_{\nu_1\cdots\nu_p}\,,
\end{equation}
where
\begin{equation}
A^{\mu_1\cdots\mu_p\nu_1\cdots\nu_p}=
\Alt_{\mu_1\cdots\mu_p}\Alt_{\nu_1\cdots\nu_p} a^{\mu_1\nu_1}\cdots
a^{\mu_p\nu_p}
\end{equation}
and
\begin{equation}
(B\varphi)_{\mu_1\cdots\mu_p} =B_{\mu_1\cdots\mu_p\nu_1\cdots\nu_p}
\varphi^{\nu_1\cdots\nu_p}\,,
\end{equation}
where
\begin{equation}
B_{\mu_1\cdots\mu_p\nu_1\cdots\nu_p}
=\Alt_{\mu_1\cdots\mu_p}\Alt_{\nu_1\cdots\nu_p} b_{\mu_1\nu_1}\cdots
b_{\mu_p\nu_p}
\end{equation}
Here $\Alt_{\mu_1\cdots\mu_p}$ denotes the complete antisymmetrization
over the indices \linebreak
$\mu_1,\dots,\mu_p$.

We will assume that these maps are isomorphisms as well. Strictly
speaking, one has to prove this.  This is certainly true for the
weakly deformed maps (maps close to the identity).
Then the inverse operator
\begin{equation}
A^{-1}: \Lambda^p\to \Lambda_{p}\,,
\end{equation}
is defined by
\begin{equation}
(A^{-1}\varphi)_{\mu_1\cdots\mu_p}
=(A^{-1})_{\mu_1\cdots\mu_p\nu_1\cdots\nu_p}
\varphi^{\nu_1\cdots\nu_p}\,,
\end{equation}
where $A^{-1}$ is determined by the equation
\begin{equation}
(A^{-1})_{\mu_1\cdots\mu_p\nu_1\cdots\nu_p}
A^{\nu_1\cdots\nu_p\alpha_1\cdots\alpha_p}
=\delta^{\alpha_1}_{[\mu_1}\cdots\delta^{\alpha_p}_{\mu_p]}\,.
\end{equation}
Notice that because of the noncommutativity, the inverse operator
$A^{-1}$ is not equal to the operator $B$, so that $A^{-1}B\ne\Id$.

This is used further to define the natural inner product on the
space of $p$-forms $\Lambda_p$ via
\begin{equation}
\left\langle\psi,\varphi\right\rangle ={1\over
  p!}\bar\psi_{\mu_1\cdots\mu_p} A^{\mu_1\cdots\mu_p\nu_1\cdots\nu_p}
\varphi_{\nu_1\cdots\nu_p}\,.
\end{equation}


\subsection{Non-commutative Star Operators}
\label{section2.5}

Of course, (on orientable manifolds) we always have the standard volume
form $\varepsilon$, which is a tensor from $E_n$ given by the completely
antisymmetric Levi-Civita symbol $\varepsilon_{\mu_1\cdots\mu_n}$.  The
contravariant Levi-Civita symbol $\tilde\varepsilon$ with components
\begin{equation}
\varepsilon^{\mu_1\cdots\mu_n}=\varepsilon_{\mu_1\cdots\mu_n}\,,
\end{equation}
is a tensor from $E^n$.
These forms are used to define the standard isomorphisms
\begin{equation}
\varepsilon: \Lambda^p\to \Lambda_{n-p}\,,\qquad
\tilde\varepsilon: \Lambda_p\to \Lambda^{n-p}
\end{equation}
by
\begin{equation}
(\varepsilon\varphi)_{\mu_1\cdots\mu_{n-p}} ={1\over p!}
\varepsilon_{\mu_1\cdots\mu_{n-p}\nu_1\cdots\nu_p}
\varphi^{\nu_1\cdots\nu_p}\,, \quad
(\tilde\varepsilon\varphi)^{\mu_1\cdots\mu_{n-p}} ={1\over p!}
\varepsilon^{\mu_1\cdots\mu_{n-p}\nu_1\cdots\nu_p}
\varphi_{\nu_1\cdots\nu_p}\,.
\end{equation}
By using the well known identity
\begin{equation}
\varepsilon_{\mu_1\cdots\mu_{n-p}\nu_1\cdots\nu_p}
\varepsilon^{\mu_1\cdots\mu_{n-p}\lambda_1\cdots\lambda_p} 
=(n-p)!p!  \delta^{\lambda_1}_{[\nu_1}\cdots\delta^{\lambda_p}_{\nu_p]}
\label{40}
\end{equation}
we get
\begin{equation}
\tilde\varepsilon\varepsilon =\varepsilon\tilde\varepsilon 
=(-1)^{p(n-p)}\Id\,.
\end{equation}

By combining $\varepsilon$ and $\tilde\varepsilon$ with the
endomorphism $\rho$ we get the forms $\varepsilon\rho^2
\in E_n$  and $\tilde\varepsilon\rho^{-2}\in E^n$.
Notice, however, that, in general, the contravariant form
$\tilde\varepsilon\rho^{-2}$ is not equal to that obtained by raising
indices of the covariant form $\varepsilon\rho^2$, i.e.
$\tilde\varepsilon\rho^{-2}\ne A\varepsilon\rho^2$ or
\begin{equation}
\varepsilon^{\mu_1\cdots\mu_n}\rho^{-2} \ne
A^{\mu_1\cdots\mu_n\nu_1\cdots\nu_n}
\varepsilon_{\nu_1\cdots\nu_n}\rho^2\,.
\end{equation}
If we require this to be the case then the matrix $\rho$ should be
defined by
\begin{equation}
\rho=\eta^{-1/4}\,,
\end{equation}
where
\begin{equation}
\eta={1\over n!}\varepsilon_{\mu_1\cdots\mu_n}
\varepsilon_{\nu_1\cdots\nu_n}a^{\mu_1\nu_1}\cdots a^{\mu_n\nu_n}\,.
\end{equation}
Since $a^{\mu\nu}$ is
self-adjoint, we also find that $\eta$ and, hence, $\rho$ is
self-adjoint.  The problem is that in general $\eta$ is not positive
definite.  Notice that in the commutative limit
\begin{equation}
\eta=\det g^{\mu\nu}=(\det g_{\mu\nu})^{-1}\,,
\end{equation}
which is strictly positive.

Therefore, we can finally define two different star operators
\begin{equation}
*, \tilde *: \Lambda_p\to \Lambda_{n-p}
\end{equation}
by
\begin{equation}
*=\varepsilon\rho A\rho\,,\qquad \tilde
*=\rho^{-1}A^{-1}\rho^{-1}\tilde\varepsilon
\end{equation}
that is
\begin{eqnarray}
(*\varphi)_{\mu_1\cdots\mu_{n-p}} &=&{1\over
  p!}\varepsilon_{\mu_1\cdots\mu_{n-p}\nu_1\cdots\nu_p} \rho
A^{\nu_1\cdots\nu_p\alpha_1\cdots\alpha_p}\rho
\varphi_{\alpha_1\cdots\alpha_p}\,,
\\
(\tilde *\varphi)_{\mu_1\cdots\mu_{n-p}}
&=&{1\over p!}\rho^{-1}
(A^{-1})_{\mu_1\cdots\mu_{n-p}\beta_1\cdots\beta_{n-p}}
\rho^{-1}
\varepsilon^{\beta_1\cdots\beta_{n-p}\alpha_1\cdots\alpha_p}
\varphi_{\alpha_1\cdots\alpha_p}\,.
\end{eqnarray}
The star operators are self-adjoint in the sense
\begin{equation}
\left\langle\varphi,*\psi\right\rangle
=\left\langle*\varphi,\psi\right\rangle\,,\qquad
\left\langle\varphi,\tilde *\psi\right\rangle=\left\langle\tilde
*\varphi,\psi\right\rangle,
\end{equation}
and satisfy the relation: for any $p$ form
\begin{equation}
*\tilde *=\tilde * * = (-1)^{p(n-p)}\Id\,.
\end{equation}

\subsection{Finsler geometry}\label{section2.4}

The above construction is closely related to Finsler
geometry~\cite{rund59}. 
Let $h(\xi)$ be an eigenvalue of the matrix 
$H(\xi)=a^{\mu\nu}\xi_\mu\xi_\nu$.
First of all, we note that 
$h(\xi)$ is a homogeneous function of $\xi$ of degree $2$,
i.e.\ for any $\lambda>0$
\begin{equation}
h(\lambda\xi)=\lambda^2 h(\xi)\,.
\end{equation}
Next, for each eigenvalue $h(\xi)$ we define the Finsler metric
\begin{equation}
g^{\mu\nu}(\xi)={1\over 2} {\partial^2
\over\partial\xi_\mu\partial\xi_\nu }h(\xi)\,.
\end{equation}
All these metrics are positive definite.
In the case when a
Finsler metric does not depend on $\xi$ it is simply a Riemannian
metric.
The Finsler metrics are homogeneous functions of $\xi$ of degree $0$
\begin{equation}
g^{\mu\nu}(\lambda\xi)=g^{\mu\nu}(\xi)\,,
\end{equation}
so that they depend only on the direction of the covector $\xi$ but
not on its magnitude.  This leads to a number of identities, in
particular,
\begin{equation}
h(\xi)=g^{\mu\nu}(\xi)\xi_\mu\xi_\nu\,
\end{equation}
and
\begin{equation}
{\partial\over\partial\xi_\mu} h(\xi)
=2g^{\mu\nu}(\xi)\xi_\nu\,.
\end{equation}

Next, again for each eigenvalue we define  the tangent vector $u\in
T_xM$  by
\begin{equation}
u^\mu(\xi)=g^{\mu\nu}(\xi)\xi_\nu\,,
\end{equation}
and the inverse (covariant) Finsler metric by
\begin{equation}
g_{\mu\nu}(u(\xi))g^{\nu\alpha}(\xi)=\delta^\alpha_\mu\,,
\end{equation} 
so that
\begin{equation}
\xi_\mu=g_{\mu\nu}(u(\xi))u^\nu(\xi)\,.
\end{equation}
The existense of Finsler metrics allows one to define various 
connections, curvatures etc (for details see~\cite{rund59}).

\subsection{Vector Bundles}

Now, we assume that the manifold $M$ admits the promotion of all vector
spaces introduced locally above to smooth vector bundles over the
manifold $M$. We use script letters to distinguish the vector bunles
from the vector spaces. Moreover, we can slightly generalize the setup
and introduce vector bundles of densities of weight $w$  over the
manifold $M$. For each bundle we indicate the weight explicitly in the
notation of the vector bundle. For example, ${\cal S}[w]$ is a vector
bundle of densities of weight $w$ with the typical fiber $S$. The
sections $\varphi$ of the vector bundle ${\cal S}[w]$ are vector-valued
functions $\varphi(x)$ that transform under diffeomorphisms
$x'^\mu=x'^\mu(x)$ according to
\begin{equation}
\varphi'(x')=J^{-w}(x)\varphi(x)\,,
\end{equation}
where
\begin{equation}
J(x)=\det\left[{\partial x'^\mu(x)\over \partial x^\alpha}\right]\,.
\end{equation}

We will consider mostly the bundles of densities of weight ${1\over 2}$,
${\cal S}[{1\over 2}]$, and, more generally, $\Lambda_p[{1\over 2}]$.
If $dx=dx^1\wedge \cdots \wedge dx^n$ is the standard Lebesgue measure
in a local chart on $M$, then we define the
diffeomorphism-invariant $L^2$-inner product
\begin{equation}
(\psi,\varphi)=\int\limits_M dx\,\left\langle\psi(x),\varphi(x)\right\rangle\,,
\end{equation}
and the $L^2$ norm
\begin{equation}
||\varphi||^2=(\varphi,\varphi)= \int\limits_M
dx\,\left\langle\varphi(x),\varphi(x)\right\rangle.
\end{equation}
The completion of $C^\infty(\Lambda_p[{1\over 2}])$ in this norm
defines the Hilbert space $L^2(\Lambda_p[{1\over 2}])$.


To avoid misunderstanding we stress here the weights of the objects
introduced above. The matrices $\Gamma^\mu$ and $a^{\mu\nu}$ have
weight $0$ and the matrix $\rho$ is assumed to be a density of weight 
${1\over 2}$. The square of this matrix, $\rho^2$, has weight $1$
and plays the role of a ``non-commutative measure''.

The operators $\varepsilon$ and $\tilde\varepsilon$ introduced above
change the weight by $1$. The operator $\varepsilon$ raises the weight by $1$,
and the operator $\tilde\varepsilon$ lowers the weight by $1$.
More precisely, for any $w$
\bea
&&\varepsilon: \Lambda^p[w]\to\Lambda_{n-p}[w-1]
\\[10pt]
&&\tilde\varepsilon: \Lambda_p[w]\to\Lambda^{n-p}[w+1]
\eea
The star operators $*$ and $\tilde *$ do not change the weights,
however,
\be
*,\ \tilde *: \Lambda_p[w]\to\Lambda_{p}[w]\,.
\ee
This is precisely the reason for the introduction of the matrix $\rho$,
which is a density of weight ${1\over 2}$.

Our goal is to construct covariant self-adjoint first-order and
second-order
differential operators acting on smooth sections of the bundles
$\Lambda_p[{1\over 2}]$ and $\Lambda^p[{1\over 2}]$, that are
covariant under both diffeomorphisms,
\begin{equation}
L'\varphi'(x')=J^{-1/2}(x)L\varphi(x)\,,
\end{equation}
and the gauge transformations
\begin{equation}
L'\varphi'=UL\varphi\,.
\end{equation}

\subsection{Non-commutative Exterior Calculus}\label{section3.1}

Next, we define invariant differential operators on smooth
sections of the bundles $\Lambda_p[0]$ and $\Lambda^p[1]$.  The
exterior derivative (the {gradient}) on tensors
\begin{equation}
d: C^\infty(\Lambda_p[0])\to C^\infty(\Lambda_{p+1}[0])\,
\end{equation}
is defined by
\begin{eqnarray}
(d \varphi)_{\mu_1\cdots\mu_{p+1}}
&=&(p+1)\partial_{[\mu_1}\varphi_{\mu_2\cdots\mu_p]}\,, 
\qquad
\mathrm{if}\ p=0,\dots,n-1\,,
\\[10pt]
d\varphi&=&0 \qquad\qquad\qquad\qquad\qquad 
\mathrm{if} \ p=n\,,
\end{eqnarray}
where the square brackets denote the complete antisymmetrization.  The
coderivative (the {divergence}) on densities of weight $1$
\begin{equation}
\tilde d: C^\infty(\Lambda^p[1])\to C^\infty(\Lambda^{p-1}[1])\,
\end{equation}
is defined by
\begin{equation}
\tilde d=(-1)^{np+1}\tilde\varepsilon d\varepsilon\,.
\label{div}
\end{equation}
By using~(\ref{40}) one can easily find
\begin{eqnarray}
(\tilde d \varphi)^{\mu_1\cdots\mu_{p-1}}
&=&\partial_\mu \varphi^{\mu\mu_1\cdots\mu_{p-1}}\,
\qquad \hbox{if} \ p=1,\dots,n\,,
\\[10pt]
\tilde d\varphi&=&0 \qquad\qquad\qquad\ \ 
\hbox{if} \ p=0\,.
\end{eqnarray}
One can also show that these definitions are covariant and satisfy the
standard relations
\begin{equation}
d^2=\tilde d^2=0\,.
\end{equation}

Recall that  the endomorphism $\rho$ is a section of the bundle
$\End({\cal S})[{1\over 2}]$.  Therefore, if $\varphi$ is a section of
the bundle $\Lambda_p[{1\over 2}]$, the quantity $\rho^{-1}\varphi$ is a 
section of the bundle $\Lambda_p[0]$.  Hence, the derivative
$d(\rho^{-1}\varphi)$ is  well defined as a smooth section of the vector
bundle $\Lambda_{p+1}[0]$.  By scaling back  with the factor $\rho$ we
get an invariant differential operator on densities of weight ${1\over 2}$
\begin{equation}
\rho d\rho^{-1}: \ 
C^\infty\left(\Lambda_p \left[{\textstyle{1\over 2}}\right]\right)
\to C^\infty\left(\Lambda_{p+1}\left[{\textstyle{1\over 2}}\right]\right).
\end{equation}
Similarly, we can define the invariant operator of codifferentiation
on densities of weight ${1\over 2}$
\begin{equation}
\rho^{-1}\tilde d\rho: \ C^\infty\left(\Lambda^p
\left[{\textstyle{1\over 2}}\right]\right)\to
C^\infty\left(\Lambda^{p-1} 
\left[{\textstyle{1\over 2}}\right]\right).
\end{equation}


\subsection{Non-commutative Connection}

Now, let ${\cal B}$ be a smooth anti-self-adjoint section of the
vector bundle $E_1[0]$, defined by the matrix-valued covector ${\cal
  B}_\mu$,
that transforms under the gauge transformations as
\begin{equation}
{\cal B}'_\mu = U{\cal B}_\mu U^{-1} -(\partial_\mu
U)U^{-1}\,.
\end{equation}
Such a section naturally defines the maps:
\begin{equation}
{\cal B}: \Lambda_p\left[\textstyle{1\over 2}\right]\to 
\Lambda_{p+1}\left[\textstyle{1\over 2}\right]
\end{equation}
by
\begin{equation}
({\cal B}\varphi)_{\mu_1\cdots\mu_{p+1}}=(p+1){\cal B}_{[\mu_1}
  \varphi_{\mu_2\cdots\mu_{p+1}]}
\end{equation}
and
\begin{equation}
\tilde{\cal B}: \Lambda^p\left[\textstyle{1\over 2}\right]\to
\Lambda^{p-1}\left[\textstyle{1\over 2}\right]
\end{equation}
by
\begin{equation}
(\tilde{\cal B}\varphi)^{\mu_1\cdots\mu_{p-1}} ={\cal
  B}_\mu\varphi^{\mu\mu_1\cdots\mu_{p-1}}\,.
\end{equation}
Notice that
\begin{equation}
\tilde{\cal B}=(-1)^{np+1}\tilde\varepsilon{\cal B}\varepsilon\,
\end{equation}
similar to~(\ref{div}).

This enables us to define the covariant exterior derivative
\begin{equation}
{\cal D}: C^\infty\left(\Lambda_p \left[{\textstyle{1\over
      2}}\right]\right)\to C^\infty\left(\Lambda_{p+1}
\left[{\textstyle{1\over 2}}\right]\right).
\end{equation}
by
\begin{equation}
{\cal D}=\rho(d+{\cal B})\rho^{-1}
\end{equation}
and the covariant coderivative
\begin{equation}
\tilde{\cal D}: C^\infty\left(\Lambda^p \left[{\textstyle{1\over
      2}}\right]\right)\to C^\infty\left(\Lambda^{p-1}
\left[{\textstyle{1\over 2}}\right]\right),
\end{equation} 
by
\begin{equation}
\tilde{\cal D}=(-1)^{np+1}\tilde\varepsilon {\cal D}\varepsilon
=\rho^{-1}(\tilde d+\tilde{\cal B})\rho \,.
\end{equation}
These operators transform covariantly under both the diffeomorphisms
and the gauge transformations.


One can easily show that the square of the operators ${\cal D}$ and
$\tilde{\cal D}$
\begin{eqnarray}
{\cal D}^2&:& 
C^\infty\left(\Lambda_p\left[{\textstyle{1\over 2}}\right]\right)
\to C^\infty\left(\Lambda_{p+2}\left[{\textstyle{1\over 2}}\right]\right)
\\
\tilde{\cal D}^2&:& 
C^\infty\left(\Lambda^{p+2}\left[{\textstyle{1\over 2}}\right]\right)
\to C^\infty\left(\Lambda^p\left[{\textstyle{1\over 2}}\right]\right)
\end{eqnarray}
are zero-order differential operators. In particular, in the case
$p=0$ they define the gauge curvature ${\cal R}$, which is a section
of the bundle $E_2[0]$, by
\begin{equation}
({\cal D}^2\varphi)_{\mu\nu} =\rho{\cal
  R}_{\mu\nu}\rho^{-1}\varphi\,,\qquad \tilde{\cal D}^2\varphi
=\rho^{-1}{\cal R}_{\mu\nu}\rho\varphi^{\nu\mu}\,,
\end{equation}
where
\begin{equation}
{\cal R}=d{\cal B}+[{\cal B},{\cal B}]\,,
\end{equation}
and the brackets $[\, ,\, ]$ denote the Lie bracket of two
matrix-valued $1$-forms, i.e.
\begin{equation}
[A,B]_{\mu\nu}=A_\mu B_\nu-B_\nu A_\mu\,.
\end{equation}

The gauge curvature is anti-self-adjoint
and transforms covariantly the gauge transformations
\begin{equation}
{\cal R}'_{\mu\nu}=U{\cal R}_{\mu\nu} U^{-1}\,.
\end{equation}

\subsection{Non-commutative Laplacians}\label{section3.3}

Finally, by using the objects introduced above we can define
second-order differential operators that are covariant under both
diffeomorphisms, and the gauge transformations.  In order to do that
we need first-order differential operators (divergences)
\begin{equation}
\mathrm{Div}: \ C^\infty\left(\Lambda_p \left[{\textstyle{1\over
      2}}\right]\right)\to C^\infty\left(\Lambda_{p-1} 
\left[{\textstyle{1\over
      2}}\right]\right),
\end{equation}

First of all, by using the $L^2$ inner product on the bundle
$\Lambda_p[{1\over 2}]$ 
we define the adjoint operator $\bar{\cal D}$ by
\begin{equation}
\left(\varphi,{\cal D}\psi\right) =\left(\bar{\cal
  D}\varphi,\psi\right).
\end{equation}
This gives
\begin{equation}
\bar{\cal D}=-A^{-1}\tilde {\cal D}A
=-(-1)^{np+1}A^{-1}\tilde\varepsilon{\cal D}\varepsilon A
=-A^{-1}\rho^{-1}(\tilde d+\tilde{\cal B})\rho A\,,
\end{equation}
which in local coordinates reads
\begin{equation}
(\bar{\cal D}\varphi)_{\mu_1\cdots\mu_p}
=-(A^{-1})_{\mu_1\cdots\mu_p\nu_1\cdots\nu_p}
\rho^{-1}(\partial_\nu+{\cal B}_\nu)\rho
A^{\nu\nu_1\cdots\nu_p\alpha_1\cdots\alpha_{p+1}}
\varphi_{\alpha_1\cdots\alpha_{p+1}}\,.
\end{equation}
The problem with this definition is that usually it is difficult to
find the matrix 
\linebreak 
$(A^{-1})_{\mu_1\cdots\mu_p\nu_1\cdots\nu_p}$.

Then we define the second order operators
\begin{equation}
\bar{\cal D}{\cal D}\,,{\cal D}\bar{\cal D}, \Delta:\
C^\infty\left(\Lambda_p \left[{\textstyle{1\over 2}}\right]\right)\to
C^\infty\left(\Lambda_{p} \left[{\textstyle{1\over 2}}\right]\right),
\end{equation}
where the ``non-commutative Laplacian'' is a self-adjoint
operator defined by
\begin{eqnarray}
\Delta&=&
-\bar{\cal D}{\cal D}-{\cal D}\bar{\cal D}
\\
&=&A^{-1}\tilde{\cal D}A{\cal D}+{\cal D}A^{-1}\tilde{\cal D}A
\nonumber\\
&=& A^{-1}\rho^{-1}(\tilde d+\tilde{\cal B})\rho A\rho(d+{\cal
  B})\rho^{-1} +\rho(d+{\cal B})\rho^{-1}A^{-1}\rho^{-1} (\tilde
d+\tilde{\cal B})\rho A\,.
\nonumber
\end{eqnarray}
In local coordinates this reads
\begin{eqnarray}
&&(\Delta\varphi)_{\mu_1\cdots\mu_p}=
\\
&&\Biggl\{(p+1)A^{-1}_{\mu_1\dots\mu_p\nu_1\dots\nu_p}
\rho^{-1}(\partial_\nu+{\cal B}_\nu)\rho
A^{\nu\nu_1\dots\nu_p\alpha\alpha_1\dots\alpha_p}
\rho(\partial_{\alpha}+{\cal B}_{\alpha})\rho^{-1}
\nonumber\\
&&
+\rho(\partial_{[\mu_1}+{\cal B}_{[\mu_1})\rho^{-1}
A^{-1}_{\mu_2\dots\mu_{p-1}]\nu_1\dots\nu_{p-1}}
\rho^{-1} (\partial_\nu+{\cal B}_\nu)\rho
A^{\nu\nu_1\dots\nu_{p-1}\alpha_1\dots\alpha_p} \Biggr\}
\varphi_{\alpha_1\dots\alpha_p}\,.
\nonumber
\end{eqnarray}

In the special case $p=0$ the ``non-commutative Laplacian" $\Delta$
reads
\begin{equation}
\Delta=\rho^{-1}(\tilde d+\tilde{\cal B})\rho A\rho(d+{\cal
  B})\rho^{-1}\,,
\end{equation}
which in local coordinates has the form
\begin{equation}
\Delta=\rho^{-1}(\partial_\mu+{\cal B}_\mu)\rho
a^{\mu\nu}\rho(\partial_\nu+{\cal B}_\nu)\rho^{-1}\,.
\end{equation}
The leading symbol of the operator $(-\Delta)$ for $p=0$
\be
\sigma_L(-\Delta;x,\xi)=a^{\mu\nu}(x)\xi_\mu\xi_\nu\,,
\ee
is self-adjoint and positive definite for $\xi\ne 0$. Therefore, the
Laplacian is an elliptic operator. The same is true for any $p$.

We could have also defined the coderivatives by
\begin{equation}
\bar{\cal D}_1=-* {\cal D}* \,,\qquad \bar{\cal D}_2=-B\tilde{\cal D}A
\,,\qquad \bar{\cal D}_3=-\tilde * {\cal D}* \,,\qquad \bar{\cal
  D}_4=-*{\cal D}\tilde *\,.
\end{equation}
These operators have the advantage that $\bar{\cal D}_1$ is polynomial
in the matrix $a^{\mu\nu}$ and $\bar{\cal D}_2$ is polynomial in the
matrices $a^{\mu\nu}$ and $b_{\mu\nu}$.  However, the second order
operators $\bar{\cal D}_j{\cal D}$, ${\cal D}\bar{\cal D}_j$ and
$\Delta_j=-\bar{\cal D}_j{\cal D}-{\cal D}\bar{\cal D}_j$,
$(j=1,2,3,4)$, are not self-adjoint, in general.  In the commutative
limit all these definitions coincide with the standard de Rham
Laplacian.

\subsection{Non-commutative Dirac Operator}\label{section3.4}

Notice first that the matrix $\Gamma$ introduced above
naturally defines a map
\begin{equation}
\Gamma:\ C^\infty(\Lambda^p\left[{\textstyle{1\over 2}}\right])
\to C^\infty(\Lambda^{p+1}\left[{\textstyle{1\over 2}}\right])
\end{equation}
by
\begin{equation}
(\Gamma\varphi)^{\mu_1\dots\mu_{p+1}}
=(p+1)\Gamma^{[\mu_1}\varphi^{\mu_2\dots\mu_{p+1}]}
\end{equation}
and the map
\begin{equation}
\tilde\Gamma:\ C^\infty(\Lambda_p\left[{\textstyle{1\over 2}}\right])
\to C^\infty(\Lambda_{p-1}\left[{\textstyle{1\over 2}}\right])
\end{equation}
by
\begin{equation}
(\tilde\Gamma\varphi)_{\mu_1\dots\mu_{p-1}}
=\Gamma^\mu\varphi_{\mu\mu_1\dots\mu_p}\,.
\end{equation}

Therefore, we can define first-order invariant differential operator
(``non-commutative Dirac operator'')
\begin{equation}
D: C^\infty\left(\Lambda_p \left[{\textstyle{1\over 2}}\right]\right)\to
C^\infty\left(\Lambda_{p} \left[{\textstyle{1\over 2}}\right]\right)\
\end{equation}
by
\begin{equation}
D=i\tilde\Gamma {\cal D} 
= i\tilde\Gamma\rho(d+{\cal B})\rho^{-1}\,,
\end{equation}
which in local coordinates reads
\begin{equation}
(D\varphi)_{\mu_1\dots\mu_p} 
= i(p+1)\Gamma^\mu
\rho(\partial_{[\mu}+{\cal
    B}_{[\mu})\rho^{-1}\varphi_{\mu_1\dots\mu_p]}\,.
\end{equation}
The adjoint of this operator with respect to the $L^2$ inner product
is
\begin{equation}
\bar D=iA^{-1}\tilde {\cal D}\Gamma A 
= iA^{-1}\rho^{-1}(\tilde
d+\tilde{\cal B})\rho\Gamma A\,,
\end{equation}
which in local coordinates becomes
\begin{equation}
(\bar D\varphi)_{\mu_1\dots\mu_p}
= i(p+1)A^{-1}_{\mu_1\dots\mu_p\nu_1\dots\nu_p}
\rho^{-1}(\partial_{\nu}+{\cal B}_{\nu})\rho
\Gamma^{[\nu}A^{\nu_1\dots\nu_p]\alpha_1\dots\alpha_p}
\varphi_{\alpha_1\dots\alpha_p}\,.
\end{equation}

In the case $p=0$ these operators simplify to
\be
D=i\tilde\Gamma {\cal D} 
=i\Gamma^\mu
\rho(\partial_{\mu}+{\cal B}_{\mu})\rho^{-1}\,,
\ee
\be
\bar D = i\tilde {\cal D}\Gamma  
= i\rho^{-1}(\partial_{\nu}+{\cal B}_{\nu})\rho\Gamma^{\nu}\,.
\ee
These operators have the same leading symbol
\be
\sigma_L(D;x,\xi)=\sigma_L(\bar D;x,\xi)=-\Gamma^\mu(x)\xi_\mu\,,
\ee
which is self-adjoint and non-degenerate. Therefore, the Dirac operator
and its adjoint $\bar D$ are elliptic. One can show that thew same is true
for any $p$.

These operators can be used then to define second order
differential operators
\begin{eqnarray}
D \bar D&=&-\tilde\Gamma{\cal D}A^{-1}\tilde{\cal D}\Gamma A
\nonumber\\
&=&-\tilde\Gamma\rho(d+{\cal B})\rho^{-1}A^{-1}\rho^{-1}
(\tilde d+\tilde{\cal B})\rho\Gamma A\,,
\\[10pt]
\bar D D&=&-A^{-1}\tilde{\cal D}\Gamma A \tilde\Gamma {\cal D}
\nonumber\\
&=&-A^{-1}\rho^{-1}(\tilde d+\tilde{\cal B})\rho
\Gamma A\tilde\Gamma \rho(d+{\cal B})\rho^{-1}\,.
\end{eqnarray}
The operators $D\bar D$ and $\bar DD$ are self-adjoint elliptic and
non-negative. They have the same non-zero spectrum. That is, if
$\lambda\ne 0$ is an eigenvalue of the operator $D\bar D$ with the
eigensection $\varphi$, then $\bar D\varphi$ is the eigenfunction of the 
operator $\bar DD$ with the same eigenvalue. Conversely, if $\psi$ is an
eigensection of the operator $\bar DD$ with an eigenvalue $\lambda\ne
0$, then $D\psi$ is an eigensection of the operator $D\bar D$ with the
same eigenvalue. Of course, if the Dirac operator is self-adjoint, i.e.
$D=\bar D$, then $D\bar D=\bar DD$. However, if $D$ is not
self-adjoint, then the zero eigenspaces of these operators can be
different, and one can define an index
\be
\mathrm{Ind}(D)=\dim\Ker(\bar D)-\dim \Ker(D)\,.
\ee

In the case $p=0$ these operators have the form
\begin{eqnarray}
D\bar D&=&-\tilde\Gamma{\cal D}\tilde{\cal D}\Gamma
\nonumber\\
&=&-\Gamma^\mu \rho(\partial_\mu+{\cal B}_\mu)
\rho^{-2}(\partial_\nu+{\cal B}_\nu)\rho\Gamma^\nu\,,
\\[10pt]
\bar DD&=&-\tilde{\cal D}\Gamma\tilde\Gamma{\cal D}
\nonumber\\
&=&-\rho^{-1}(\partial_\nu+{\cal B}_\nu)
\rho\Gamma^\nu \Gamma^\mu
\rho(\partial_\mu+{\cal B}_\mu)\rho^{-1}\,.
\end{eqnarray}
These operators have the same leading symbol as the non-commutative 
Laplacian. Therefore, one can obtain a non-commutative version of the
Lichnerowicz formula.

These constructions can be used to develop non-commutative
generalization of the standard theory of elliptic complexes, in
particular, spin complex, de Rham complex, index theorems,  cohomology
groups, heat kernel etc.   If the bundle ${\cal S}$ is $\ZZ_2$-graded,
then, similarly to the Riemannian case discussed in section 2, there is
an index of the Dirac operator even it is self-adjoint. This is a very
interesting topic that requires further study. 


\subsection{Spectral Asymptotics}\label{section5}

Since the
non-zero spectra of the operators $\bar DD$ and $D\bar D$ are
isomorphic, this also means that the spectral invariants of the
operators $\bar DD$ and $D\bar D$ are equal except possibly for the
invariant $A_{n/2}(\II,\bar DD)$ which determines the index in even
dimension. Thus, for $n>2$ the spectral invariants $A_0$ and $A_1$ of
the operators $\bar DD$ and $D\bar D$ are the same. Therefore, we can
pick any of these operators $\bar DD$ or $D\bar D$
to compute the invariants $A_0$ and $A_1$.
In present paper we will restrict ourselves to the case $p=0$.
The operators $\bar DD$ and $D\bar D$ 
have the same leading symbol equal to
\be
\sigma_L(\bar DD;x,\xi)
=\sigma_L(D\bar D;x,\xi)
=H(x,\xi)=[\Gamma^\mu(x)\xi_\mu]^2\,,
\ee
with $\xi\in T_x^*M$ a cotangent vector.
Since by our assumption this matrix is self-adjoint
and positive definite, these operators 
are elliptic. In fact, all non-commutative Laplacians and Dirac
operators introduced in the previous subsection are elliptic.

It is well known that a self-adjoint elliptic
partial differential operator with positive definite leading symbol on
a compact manifold without boundary has a discrete real spectrum
bounded from below~\cite{gilkey95}.  Since the operator $\bar DD$
transforms
covariantly under the diffeomorphisms as well as under the gauge
transformations~(\ref{gt0}) the spectrum is {invariant} under
these transformations.

The heat semigroup $\exp(-t\bar DD)$ is a trace-class operator with a well
defined $L^2$ trace
\begin{equation}
\Tr_{L^2}\exp(-t\bar DD) =\int\limits_M \,dx\,\tr_S U(t;x,x)\,.
\end{equation}
Moreover for any  smooth  endomorphism-valued
function $F\in C^\infty(\End({\cal S})[0])$
the following trace is defined
\begin{equation}
\Tr_{L^2}[F\exp(-t\bar DD)] =\int\limits_M \,dx\,\tr_S F(x)U(t;x,x)\,.
\end{equation}
We have defined the heat
kernel $U(t;x,x')$
in such a way that it transforms as a density of weight
${1\over 2}$ at both points $x$ and $x'$. More precisely, it is a
section of the exterior tensor product bundle ${\cal S}[{1\over
    2}]\boxtimes{\cal S}^*[{1\over 2}]$.  Therefore, the heat kernel
diagonal $U(t;x,x)$
transforms as a density of weight $1$, i.e.\ it is a section
of the bundle $\End({\cal S})[1]$, and the trace 
$\Tr_{L^2}\exp(-t\bar DD)$
 is invariant under diffeomorphisms. 

As in the case of Laplace type operators there is an asymptotic
expansion as $t\to 0^+$ of the heat kernel diagonal
\begin{equation}
U(t;x,x)\sim (4\pi)^{-n/2} \sum_{k=0}^\infty t^{(2k-n)/2}
a_k(\bar DD;x)\,,
\end{equation}
and of the heat trace as $t\to 0^+$~\cite{gilkey95}
\begin{equation}
\Tr_{L^2}[F\exp(-t\bar DD)]
\sim (4\pi)^{-n/2} \sum\limits_{k=0}^\infty t^{(2k-n)/2}
A_{k}(F,\bar DD)\,,
\label{17}
\end{equation}
where
\begin{equation}
A_k(F,\bar DD)=\int\limits_M dx\, \tr_S F(x)a_k(\bar DD;x)\,.
\end{equation}
are the global heat invariants. 

A second-order differential operator is called Laplace type if it has
a scalar leading symbol.  Most of the calculations in quantum field
theory and spectral geometry are restricted to the Laplace type
operators for which nice theory of heat kernel asymptotics is
available~\cite{gilkey95,avramidi91b,avramidi99,
  avramidi00,avramidi02}.  However, the operators condidered in the
present paper have a matrix valued principal symbol $H(x,\xi)$ and
are, therefore, not of Laplace type. The study of heat kernel
asymptotics for non-Laplace type operators is quite new and the
methodology is still underdeveloped. As a result even the
invariant $A_2$ is not known, in general. For some
partial results see~\cite{avrbrans01,avrbrans02,avramidi04b}.

\subsection{Heat Invariants}\label{section5.1}

For so called natural non-Laplace type differential operators, which are
constructed from a Riemannian metric and canonical connections on
spin-tensor bundles the coefficients $A_0$ and $A_1$ were computed
in~\cite{avrbrans02}. For general non-Laplace type operators they were
computed in \cite{avramidi04b}.  Following these papers we will use a
rather {formal} method that is sufficient for our purposes of computing
the asymptotics of the heat trace of the second-order elliptic
self-adjoint operator $\bar DD$.


First, we present the heat kernel diagonal for the operator $\bar DD$
in the form
\begin{equation}
U(t;x,x) =\int\limits_{\RR^n} {d\xi\over (2\pi)^n} e^{-i\xi
  x}\exp(-t\bar DD) e^{i\xi x} \,,
\end{equation}
where $\xi x=\xi_\mu x^\mu$, which can be transformed to
\begin{equation}
U(t;x,x)=\int\limits_{\RR^n} {d\xi\over (2\pi)^n}
\exp\left[-t\left(H+K+\bar DD\right)\right]\cdot \II\,,
\end{equation}
where $H$ is the leading symbol of the operator $\bar DD$
\begin{equation}
H=[\Gamma(\xi)]^2\,,
\end{equation}
with $\Gamma(\xi)=\Gamma^\mu(x)\xi_\mu$, and
$K$ is a first-order self-adjoint operator defined by
\be
K=-\Gamma(\xi)D-\bar D\Gamma(\xi)\,.
\ee
Here the
operators in the exponent act on the unity matrix $\II$ from the left.

By changing the integration variable $\xi\to t^{-1/2}\xi$ we obtain
\begin{equation}
U(t;x,x)=(4\pi t)^{-n/2}\int\limits_{\RR^n} {d\xi\over \pi^{n/2}} 
\exp\left(-H-\sqrt tK-t\bar DD\right)\cdot \II\,,
\end{equation}
and the problem becomes now to evaluate the first three terms of the
asymptotic expansion of this integral in powers of $t^{1/2}$ 
as $t\to 0$.  

By using the
Volterra series
\begin{eqnarray}
\exp(A+B)&=&e^A+\sum\limits_{k=1}^\infty
\int\limits_0^1 d\tau_k \int\limits_0^{\tau_k}d\tau_{k-1}\cdots 
\int\limits_0^{\tau_2} d\tau_1\times
\nonumber\\[10pt]
&&
\times\, e^{(1-\tau_{k})A} Be^{(\tau_k-\tau_{k-1}) A} \cdots
e^{(\tau_2-\tau_1)A} Be^{\tau_1 A}\,,
\end{eqnarray}
we get
\begin{eqnarray}
&&\exp\left(-H-\sqrt tK-t\bar DD\right)
=e^{-H}-
t^{1/2}\int\limits_0^1 d\tau_1 e^{-(1-\tau_1)H}K e^{-\tau_1 H}
\nonumber\\
&&
\qquad
+\,t\Biggl[ \int\limits_0^1d\tau_2\int\limits_0^{\tau_2}d\tau_1
  e^{-(1-\tau_2)H} K e^{-(\tau_2-\tau_1)H}Ke^{-\tau_1 H}-
\nonumber\\
&&
\qquad
\hphantom{+\,t\Biggl[}
  -\int\limits_0^1 d\tau_1 e^{-(1-\tau_1)H}
\bar DD e^{-\tau_1 H} \Biggr]+
O(t^2)\,.
\end{eqnarray}

Now, since $K$ is linear in $\xi$ the term proportional to $t^{1/2}$
vanishes after integration over $\xi$. Thus, we obtain the first two
coefficients of the asymptotic expansion of the
heat kernel diagonal
\begin{equation}
U(t;x,x)=(4\pi t)^{-n/2}\left[a_0(x)+ta_1(x)+O(t^2)\right]
\end{equation}
in the form
\begin{eqnarray}
a_0 &=&\int\limits_{\RR^n}{d\xi\over \pi^{n/2}}\,
e^{-H}\,,
\\
a_1&=&\int\limits_{\RR^n}{d\xi\over \pi^{n/2}}\,\Biggl[
\int\limits_0^1d\tau_2\int\limits_0^{\tau_2}d\tau_1 e^{-(1-\tau_2)H}
K e^{-(\tau_2-\tau_1)H}Ke^{-\tau_1 H}-
\nonumber\\&&
         \hphantom{\int\limits_{\RR^n}{d\xi\over \pi^{n/2}}\,\Biggl[}
-\int\limits_0^1 d\tau_1 e^{-(1-\tau_1)H}
\bar DD e^{-\tau_1 H}
\Biggr] \,.
\end{eqnarray}
These quantities are matrix-valued  densities. The coefficient $a_0$ is
constructed from the matrix $a$ but not its derivatives, whereas the
coefficient $a_1$ is constructed from the matrix $a$ and its first and
second derivatives as well as from the first derivatives of the field
${\cal B}$ and the matrix $\rho$ and its first and second derivatives.
Morevover, it is polynomial in the derivatives of $a^{\mu\nu}$, $\rho$
and ${\cal B}_\mu$, more precisely, linear in second derivatives of $a$
and $\rho$ and the first derivatives of ${\cal B}$ and quadratic in
first derivatives of $a$ and $\rho$. By tracing the local invariants and 
integrating over the maniolfd we finally get the global heat invariants
\begin{eqnarray}
A_0 &=&\int\limits_M dx\,\int\limits_{\RR^n}{d\xi\over \pi^{n/2}}\,
\tr_S\,e^{-H}\,,
\\
A_1&=&\int\limits_M dx\,\int\limits_{\RR^n}{d\xi\over \pi^{n/2}}\,
\tr_S\,\Biggl[
\int\limits_0^1d\tau_2\int\limits_0^{\tau_2}d\tau_1 e^{-(1-\tau_2)H}
K e^{-(\tau_2-\tau_1)H}Ke^{-\tau_1 H}-
\nonumber\\&&
     \hphantom{\int\limits_M dx\,\int\limits_{\RR^n}
{d\xi\over \pi^{n/2}}\,\tr_V\,\Biggl[}
-\int\limits_0^1 d\tau_1 e^{-(1-\tau_1)H}
\bar DD e^{-\tau_1 H} \Biggr] \,.
\end{eqnarray}
The global heat invariants are invariant under both the diffeomorphisms
and the gauge transformations.  Since the operator $\bar DD$ is
self-adjoint, the heat kernel diagonal $U(t;x,x)$  is a self-adjoint
matrix-valued density, and, therefore, the heat trace 
$\Tr_{L^2}\exp(-t\bar DD)$ is a real invariant. Therefore, the
coefficients $a_0$ and $a_1$ are self-adjoint matrix densities 
and the invariants $A_0$ and $A_1$ are real.


\section{Non-commutative Einstein-Hilbert Functional}

It is an
interesting fact that a linear combination of the  first two
spectral invariants of the Dirac operator on Riemannian manifold
determines the Einstein-Hilbert functional. Indeed, by using the eqs. 
(\ref{170}), (\ref{171}) we obtain 
\bea 
S_{EH}(g)&=&-{1\over 16\pi G}{1\over
N}\left\{12A_1(\II,D^2)+2\Lambda A_0(\II,D^2)\right\}
\nonumber\\[10pt]
&=&
{1\over 16\pi G}\int_M d\vol \left(R-2\Lambda\right)\,,
\eea 
where $G$ and $\Lambda$ are positive parameters. This functional is the
action functional of the general theory of relativity which determines
the  Einstein equations of the gravitational field. In the general
theory of relativity the Riemannian metric $g$ (rather its
pseudo-Riemannian version) is identified with the gravitational field
and
the parameters $G$ and $\Lambda$ with the Newtonian gravitational
constant and the cosmological constant respectively.

In differential geometry the extremals of the 
Einstein-Hilbert functional are the Einstein spaces, that is Riemannian
metrics $g$ satisfying the vacuum Einstein equations with the
cosmological constant 
\be 
R_{\mu\nu}=\Lambda g_{\mu\nu}\,, 
\ee 
where
$R_{\mu\nu}$ is the Ricci tensor. The study of Einstein spaces is a very
important area in differential  geometry and mathematical physics.

In full analogy with the above one can  build an invariant functional of
the non-commutative metric $a^{\mu\nu}$ (or the non-commutative Dirac
matrices $\Gamma^\mu$), the endomorhism  $\rho$ and the
endomorphism-valued covector ${\cal B}_\mu$  as a linear combination of
the  first two spectral invariants of the non-commutative  operator
$\bar D D$.  Such a functional can be called a non-commutative
deformation of the Einstein-Hilbert functional. The extremals of this
functional are then ``non-commutative  Einstein equations'', whose
solutions determine the structures that can be called ``non-commutative
Einstein spaces''. One can show that this functional does not depend on
the derivatives of the field ${\cal B}_\mu$. Therefore, variation with
respect to ${\cal B}_\mu$ gives  just a constraint which expresses
${\cal B}_\mu$ in terms of derivatives of the functions $a^{\mu\nu}$ and
$\rho$. One can also impose some additional consistency conditions to
express the extra ingredients, like the matrix $\rho$ in terms of the
non-commutative metric $a^{\mu\nu}$ (or non-commutative Dirac matrices
$\Gamma^\mu$). For example, the requirement that  the non-commutative
Dirac operator should be self-adjoint, gives a constraint which can be
used to fix the connection ${\cal B}$.  The question of uniqueness of
such consistency conditions is one of many open problems in this
approach. The study of these structures is an extremely interesting
problem in differential geometry. It might also have applications in
modern gravitational and high-energy physics. Such attempts are
discussed in our previous papers
\cite{avramidi03,avramidi04a,avramidi04b}.

\section*{Acknowledgements}

It was a pleasure to contribute to the special issue of the 
`International Journal of Geometric Methods in Modern Physics' 
dedicated to 100th birthday of Dmitri Ivanenko and the 75 aniversary  of
the Fock-Ivanenko coefficients. Both Dmitri Ivanenko and Vladimir Fock
were great theoreticial physicists who had a major impact on many areas
of modern theoretical and mathematical physics.


\begin{thebibliography}{999}

\bibitem{fockivanenko29a}
V. Fock and D. Ivanenko, 
{\it On a possible geometric interpretation of relativistic quantum theory},  
Z. Physik, {\bf 54} (1929) 798--802 

\bibitem{fockivanenko29b}
V. Fock and D. Ivanenko, 
{\it Quantum linear geometry and parallel displacement},
Compt. Rend. Acad. Sci. Paris, {\bf  188} (1929) 1470--1472

\bibitem{fock29a} 
V. Fock, 
{\it On Dirac equations in general relativity},
Compt. Rend. Acad. Sci. Paris, {\bf 188} (1929) 25--28

\bibitem{fock29b} 
V. Fock, 
{\it Geometrization of the Dirac electron theory}, 
Z. Physik, {\bf 57} (1929) 261--277

\bibitem{fock29c}
V. Fock, 
{\it The Dirac wave equation and the Riemann geometry},
J. Phys. Radium, {\bf 10} (1929) 392--405

\bibitem{lawson89}
H. B. Lawson and M.-L. Michelsohn, 
{\it Spin Geometry}, 
Princeton, Princeton University Press, 1989

\bibitem{berline92} 
N. Berline, E. Getzler and M. Vergne, 
{\it Heat Kernels and Dirac Operators},
Berlin, Springer-Verlag,  1992

\bibitem{friedrich00}
Th. Friedrich, 
{\it Dirac Operators in Riemannian Geometry},
Graduate Studies in Mathematics, Vol. 25,
Providence, Rhode Island, American Mathematical Society, 2000

\bibitem{avramidi03}
I. G. Avramidi,
{\it A Non-commutative Deformation of General Relativity}, 
Phys. Lett. B, {\bf 576} (2003) 195--198

\bibitem{avramidi04a}
I. G. Avramidi, 
{\it Matrix General Relativity: A new look at Old Problems}, 
Class. Quant. Grav., {\bf 21} (2004) 103--120


\bibitem{avramidi04b}
I. G. Avramidi,
{\it Gauged Gravity via Spectral Asymptotics of non-Laplace type Operators},
J. High Energy Phys., {\bf 07} (2004) 030

\bibitem{zhelnorovich82}
V. A. Zhelnorovich, 
{\it Theory of Spinors and Its Application in Physics and Mechanics},
Moscow, Nauka, 1982 [in Russian]


\bibitem{gilkey95}
P.B.~Gilkey, {\it Invariance theory, the heat equation and the
  Atiyah-Singer index theorem}, CRC Press, Boca Raton, 1995.

\bibitem{avramidi91b}
I.G.~Avramidi, {\it A covariant technique for the calculation of the
  one-loop effective action}, Nucl. Phys. B {\bf 355} (1991) {712} [Erratum
  Nucl. Phys. B {\bf 509} (1998) 557].

\bibitem{avramidi99}
I.G.~Avramidi, {\it Covariant techniques for computation of the heat
  kernel}, {Rev.\ Math.\ Phys.} {\bf 11} (1999) {947}.

\bibitem{avramidi00}
 I.G.~Avramidi, {\it Heat kernel and quantum gravity},
Springer-Verlag, Lecture Notes in Physics, Series Monographs, LNP:m64,
Berlin, New York, 2000.

\bibitem{avramidi02}
I.G.~Avramidi, {\it Heat kernel approach in quantum field theory},
Nucl. Phys. Proc. Suppl. {\bf 104} {(2002)} {3}.

\bibitem{rund59}
H.~Rund, {\it The differential geometry of Finsler spaces}, Nauka,
Moscow, 1981 (russian); Springer-Verlag, Berlin, 1959 (english).

\bibitem{avrbrans01}
I.G.~Avramidi and T.~Branson, {\it Heat kernel asymptotics of
  operators with non-Laplace principal part}, Rev. Math. Phys.
 {\bf 13} (2001) 847

\bibitem{avrbrans02}
I.G.~Avramidi and T.~Branson, {\it A discrete leading symbol and
  spectral asymptotics for natural differential operators},
{J.\ Funct.\ Analys.} {\bf 190} {(2002)} {292}.

\end{thebibliography}
\end{document}